\documentclass[12pt]{article}
\pdfoutput=1
\usepackage{a4wide,epsfig,psfrag,amsmath,amssymb,cite,scalefnt}
\usepackage[dvipsnames]{xcolor}
\usepackage{amsmath,comment,braket}
\usepackage{placeins}
\usepackage{breqn}
\usepackage{slashed}
\usepackage{subcaption}

\graphicspath{ {figs_3l/} }

\allowdisplaybreaks

\begin{document}

\title{\vskip-3cm{\baselineskip14pt
    \begin{flushleft}
      \normalsize LTH 1397, P3H-25-020, TTP25-007, ZU-TH 19/25
    \end{flushleft}} \vskip1.5cm
  Three-loop large-$N_c$ virtual corrections to $gg\to HH$ in the forward limit
}

\author{
  Joshua Davies$^{a}$,
  Kay Sch\"onwald$^{b}$,
  Matthias Steinhauser$^{c}$
  \\[1mm]
  \\
  {\small\it (a) Department of Mathematical Sciences, University of
    Liverpool,}
  {\small\it Liverpool, L69 3BX, UK}
  \\
  {\small\it (b) Physik-Institut, Universit\"at Z\"urich, Winterthurerstrasse 190,}\\
  {\small\it 8057 Z\"urich, Switzerland}
  \\
  {\small\it (c) Institut f{\"u}r Theoretische Teilchenphysik,
    Karlsruhe Institute of Technology (KIT),}\\
  {\small\it Wolfgang-Gaede Stra\ss{}e 1, 76131 Karlsruhe, Germany}
  \\
}

\date{}

\maketitle

\thispagestyle{empty}

\begin{abstract}

  We compute the three-loop form factors for $gg\to HH$ in the limit of
  vanishing transverse momentum of the Higgs boson which provides a reasonable approximation of
   the cross section.  In our calculations we adopt the
  large-$N_c$ limit, which already includes non-trivial non-planar Feynman
  diagrams.  We discuss the results for top quark masses in the pole and
  $\overline{\rm MS}$ schemes and show that the scheme dependence is
  significantly reduced at next-to-next-to-leading order.

\end{abstract}


\thispagestyle{empty}

\newpage


\section{Introduction}

Higgs boson pair production is one of the processes which will get a lot of
attention in the upcoming years -- both from the experimental and the theory
side. The main production channel is via gluon fusion which shows a sizable
dependence on the top quark mass. As a consequence, there is a relatively
strong dependence on the top quark mass renormalization scheme as has been
discussed in Refs.~\cite{Baglio:2020wgt,Bagnaschi:2023rbx}. It amounts up to  20\%
after including the next-to-leading order (NLO)
corrections into the theory predictions. In order to reduce this uncertainty
it is necessary to move to next-to-next-to-leading order (NNLO), which
involves both virtual corrections to $gg\to HH$ and real radiation
contributions with one or two additional massless partons in the final state.

\begin{figure}[b]
    \begin{center}
    \includegraphics[width=.30\textwidth]{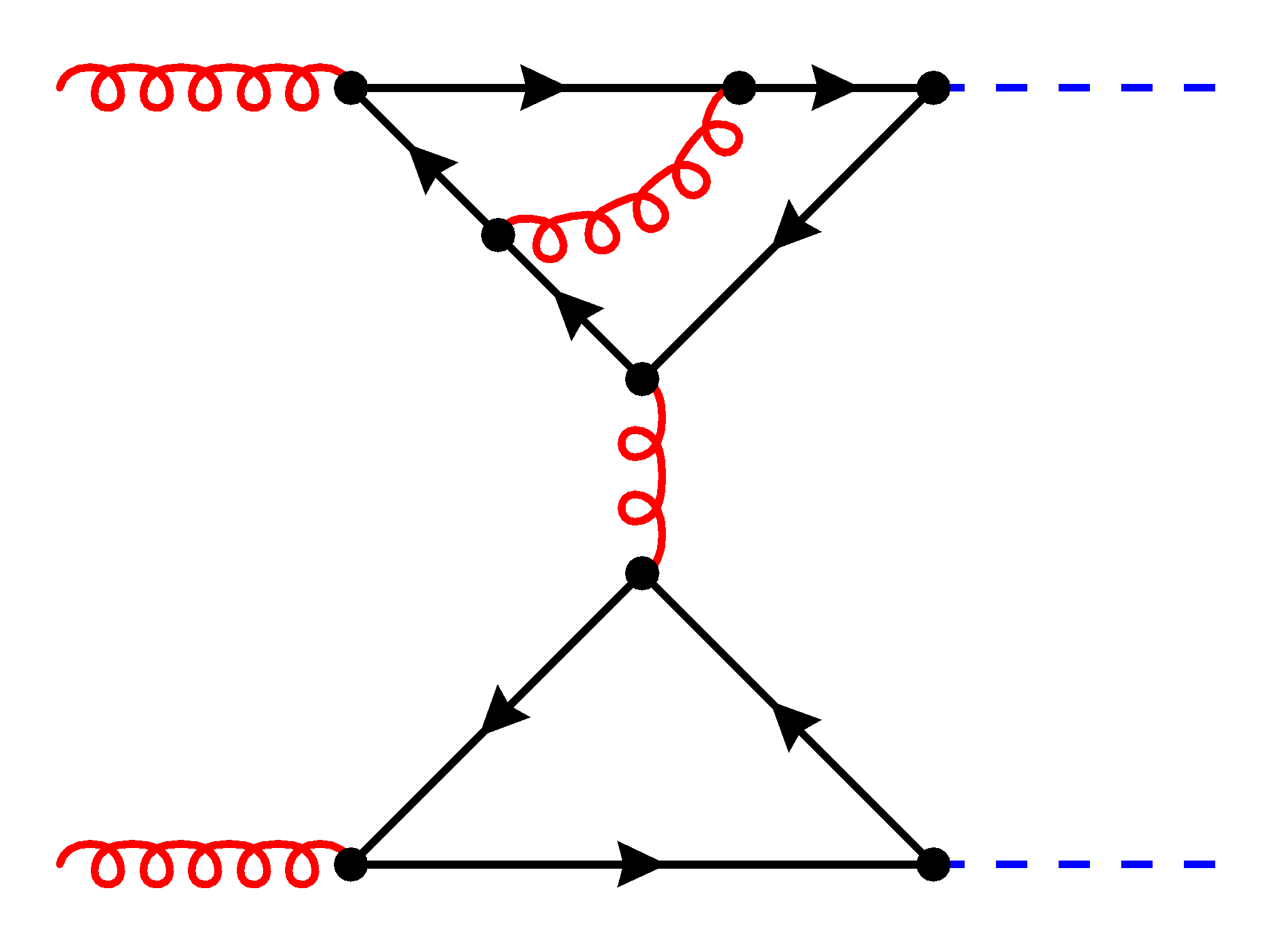}
    \includegraphics[width=.37\textwidth]{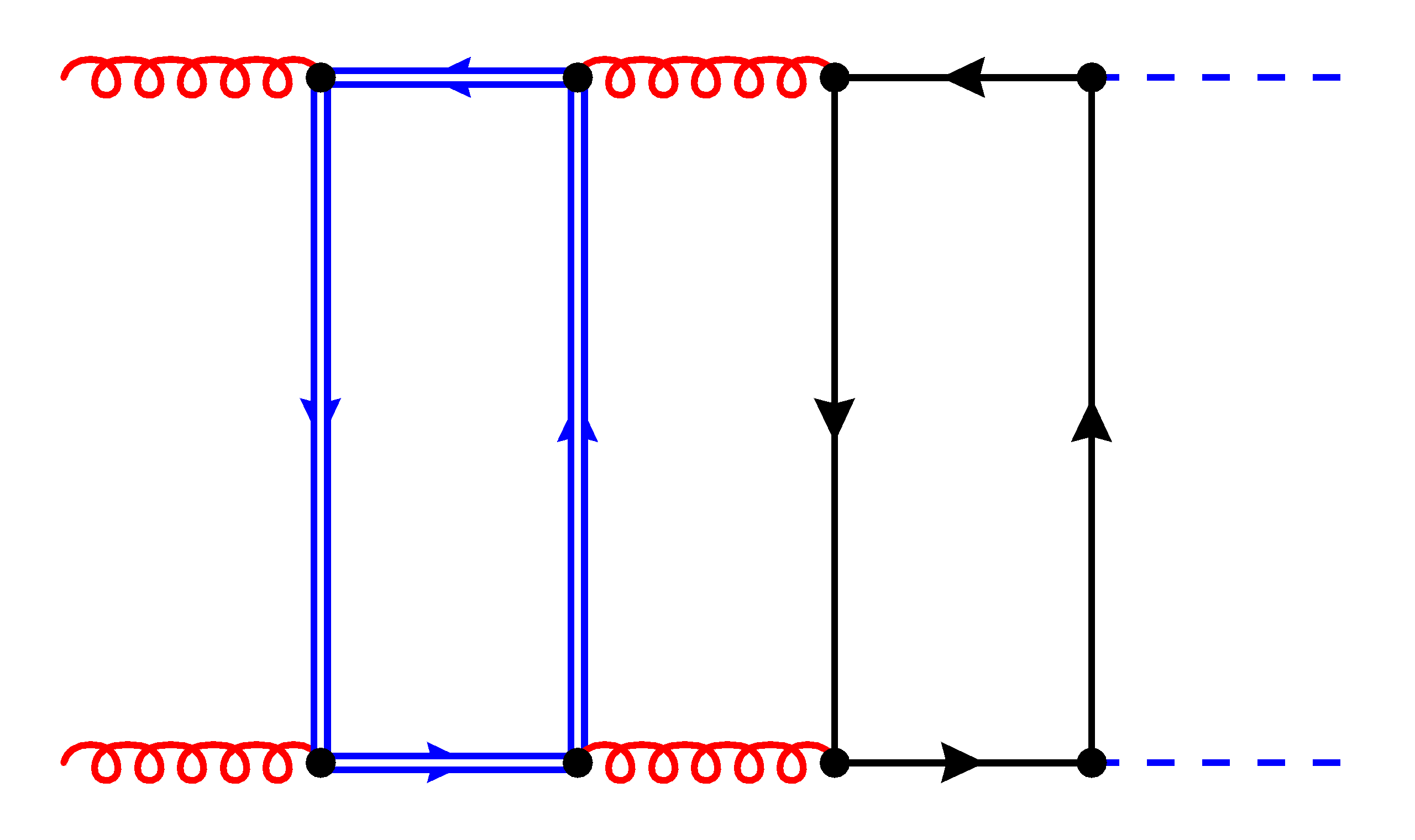}
    \includegraphics[width=.30\textwidth]{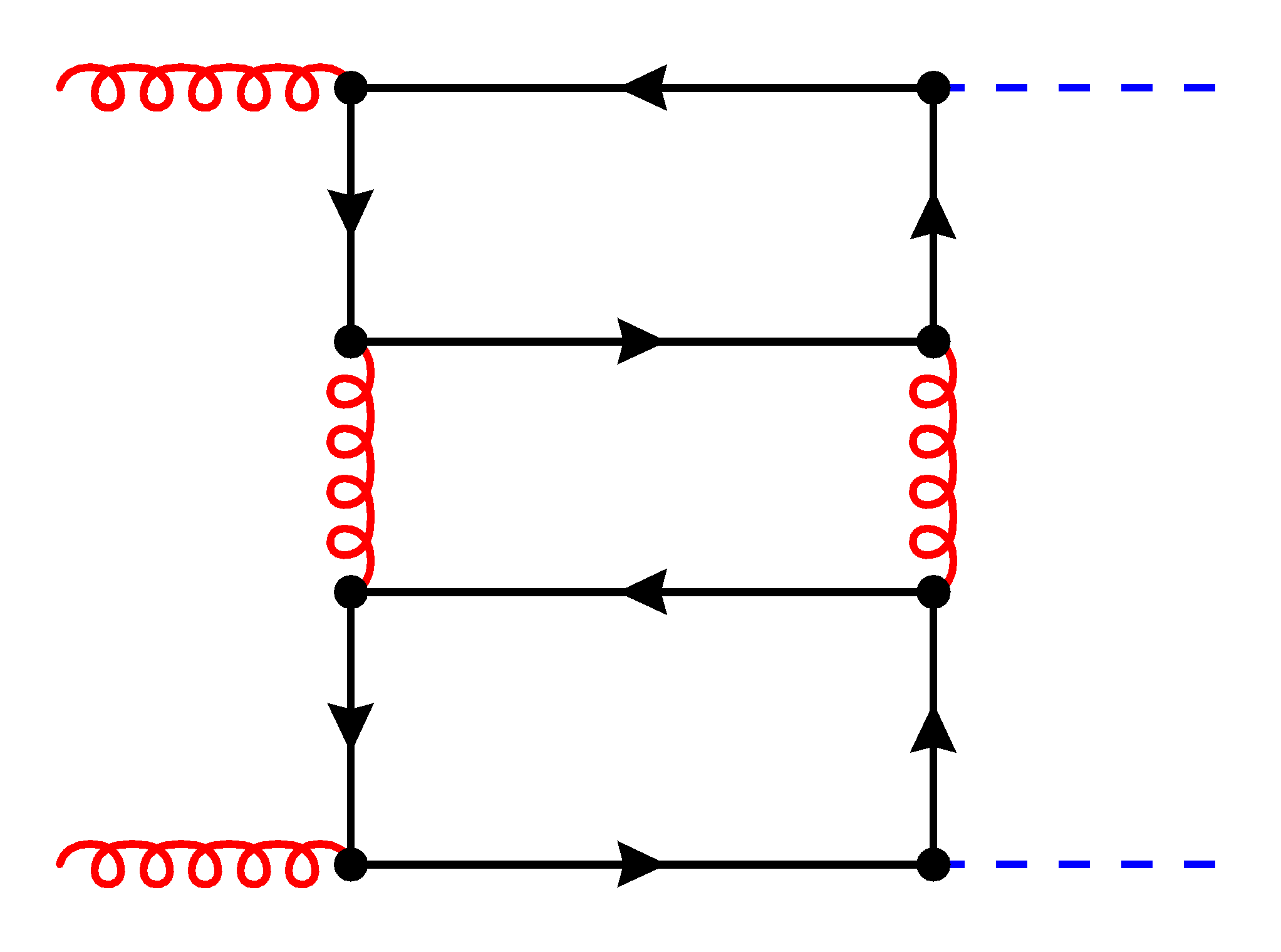}
    
    (a) 
    \hspace{0.30\textwidth}
    (b) 
    \hspace{0.30\textwidth}
    (c)
    \end{center}
  \caption{\label{fig::gghh_FDs}Sample Feynman diagrams of contributions which are not computed in the present paper. The contributions shown represent: (a) reducible contributions, (b) contributions with one closed light quark loop, 
  (c) contributions with two closed top quark loops.
  Curly (red) lines correspond to gluons,
  single (black) lines to the top quark,
  double (blue) lines to a massless quark
  and dashed (blue) lines to the Higgs boson.
  }
\end{figure}

\begin{figure}[t]
    \begin{center}
        \includegraphics[width=.24\textwidth]{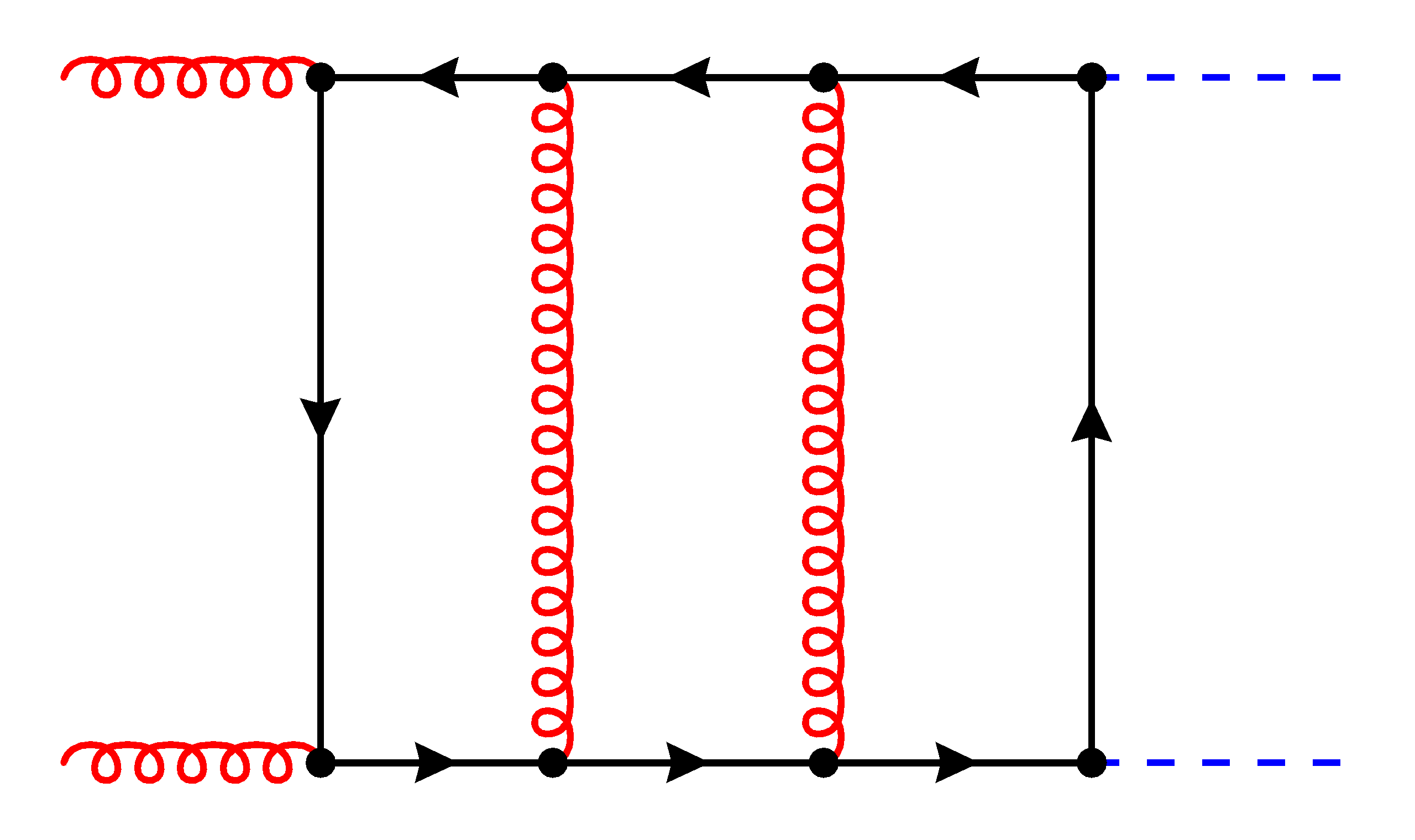}
        \includegraphics[width=.24\textwidth]{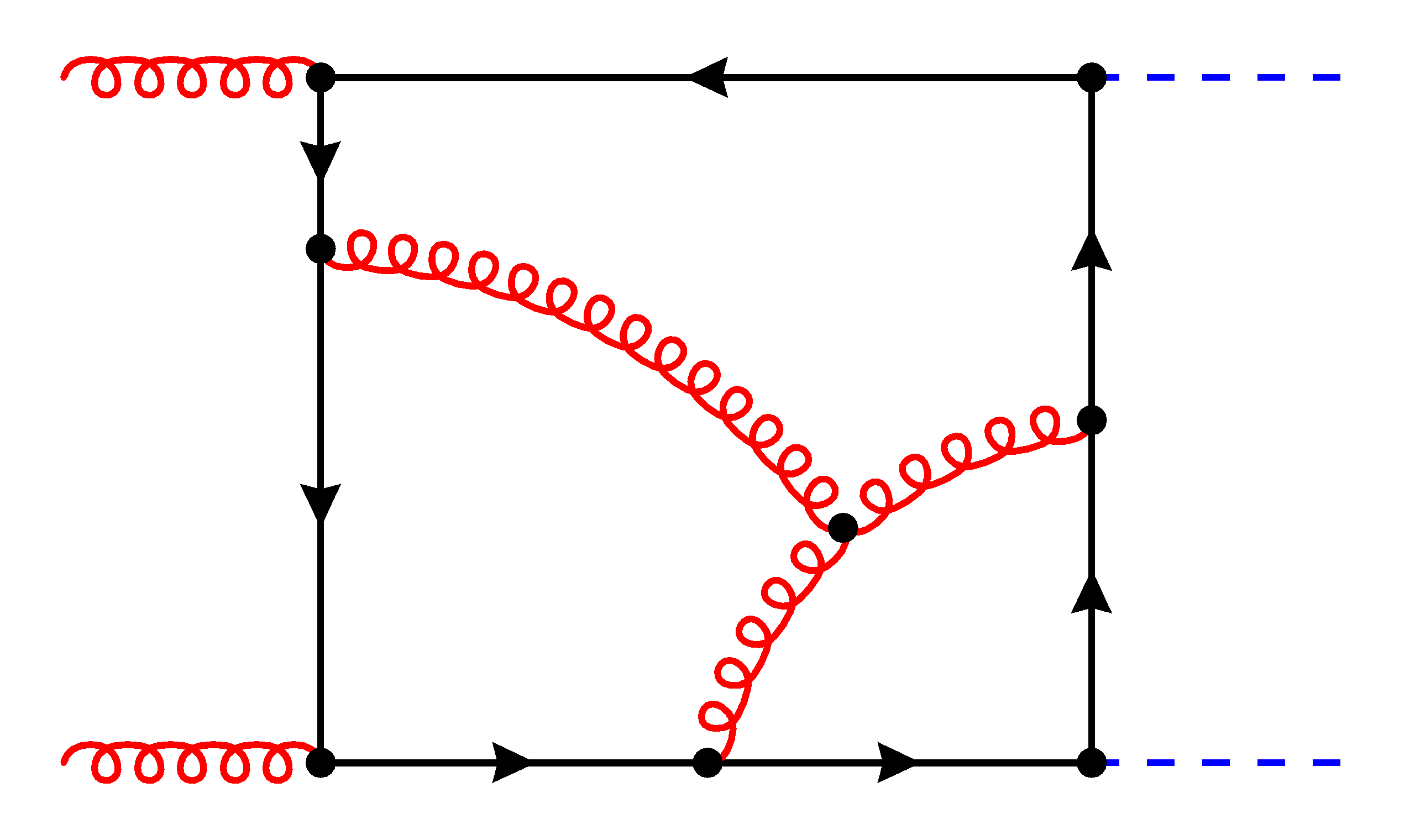}
        \includegraphics[width=.24\textwidth]{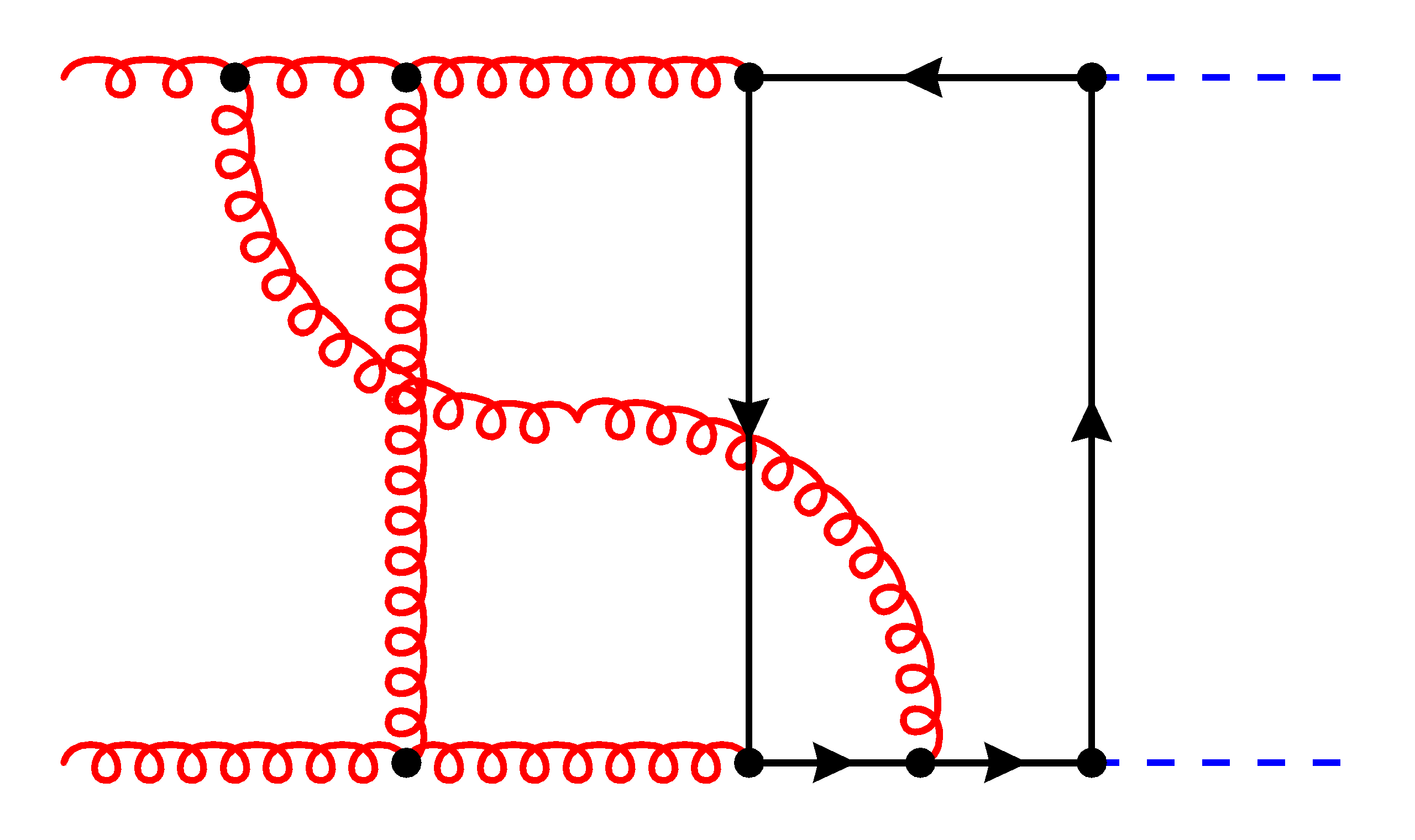}
        \includegraphics[width=.24\textwidth]{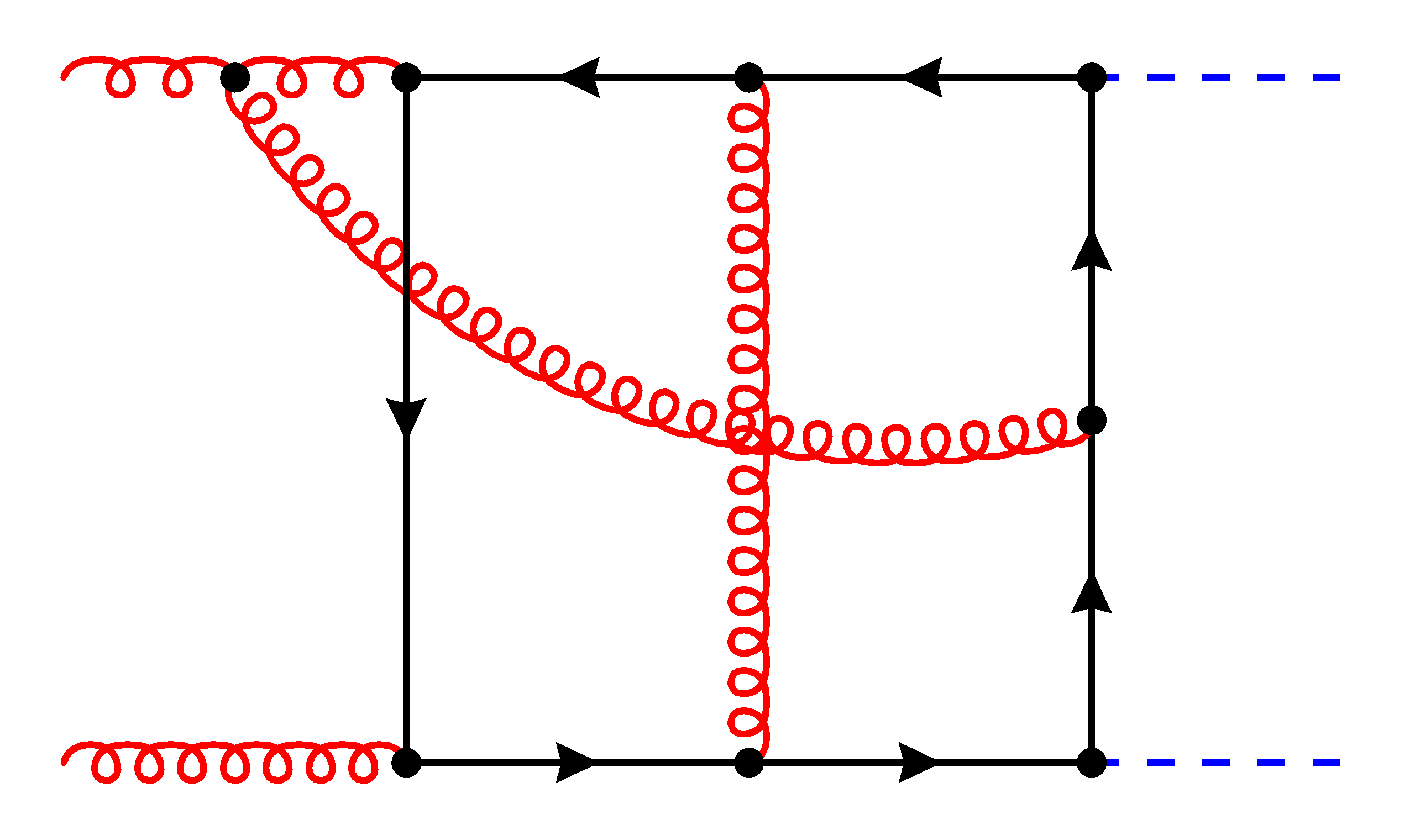}

        (a) 
        \hspace{0.2\textwidth}
        (b) 
        \hspace{0.2\textwidth}
        (c)
        \hspace{0.2\textwidth}
        (d)
    \end{center}
    \caption{\label{fig::gghh_FD2}
    Sample Feynman diagrams of the contributions containing one closed top quark loop.
    Diagrams (a)-(c) contribute in the leading-$N_c$ limit, while (d) is suppressed in this limit.
    The line styles are as in Fig.~\ref{fig::gghh_FDs}.
  }
\end{figure}

The strong scheme dependence at NLO has its origin in the two-loop virtual corrections.  
The contribution which is most important for its reduction
 is the three-loop virtual correction to the $gg\to HH$ form factors. Here $m_t$ has to be renormalized at the two-loop level.
Sample Feynman diagrams are shown in Figs.~\ref{fig::gghh_FDs} and~\ref{fig::gghh_FD2}.
In
this paper we advance the findings of Ref.~\cite{Davies:2023obx}, where light-fermion contributions have been considered (see Fig.~\ref{fig::gghh_FDs}(b)) to the next
step and compute the complete dependence on the top quark mass in the forward
limit for vanishing transverse momenta of the Higgs 
bosons, albeit restricting to the contributions with
one closed top quark loop and also 
to the large-$N_c$ (where $N_c$ stands for the number of colours in QCD) approximation, see Fig.~\ref{fig::gghh_FD2}(a) to~(c) for sample Feynman diagrams. This allows us to
estimate the reduction of the uncertainty due to the top quark mass
renormalization scheme.
One-particle reducible contributions as, e.g., shown in Fig.~\ref{fig::gghh_FDs}(a) have been computed in Ref.~\cite{Davies:2024znp} using semi-analytic methods in the whole phase space. One-particle irreducible contributions
as shown in Fig.~\ref{fig::gghh_FDs}(c) are not known beyond the large-$m_t$ limit.

We consider the process $g(q_1) + g(q_2) \to H(q_3) + H(q_4)$ where 
all momenta are incoming. Then the Mandelstam variables are given by
\begin{eqnarray}
  s=(q_1+q_2)^2\,,\qquad t=(q_1+q_3)^2\,,\qquad u=(q_2+q_3)^2\,,
\end{eqnarray}
with $s+t+u=2m_H^2$ and the transverse momentum of the Higgs boson
is given by $p_T^2 = (tu-m_H^4)/s$.
The matrix element is conveniently decomposed into two form
factors\footnote{See Ref.~\cite{Davies:2023obx} for explicit expressions for
  $A_1^{\mu\nu}$ and $A_2^{\mu\nu}$.}
\begin{eqnarray}
  {\cal M}^{ab} &=& 
  \varepsilon_{1,\mu}\varepsilon_{2,\nu}
  {\cal M}^{\mu\nu,ab}
  \,\,=\,\,
  \varepsilon_{1,\mu}\varepsilon_{2,\nu}
  \delta^{ab} X_0 s 
  \left( F_1 A_1^{\mu\nu} + F_2 A_2^{\mu\nu} \right)
  \,,
                    \label{eq::M}
\end{eqnarray}
with
\begin{eqnarray}
  X_0(\mu) &=& \frac{G_F}{\sqrt{2}} \frac{\alpha_s(\mu)}{2\pi} T_F \,,
               \label{eq::X0}
\end{eqnarray}
with the Fermi constant $G_F$, the strong coupling constant $\alpha_s(\mu)$ in 
the $\overline{\rm MS}$ scheme and $T_F=1/2$.
At higher orders we will also need the colour factors 
$C_F = (N_c^2-1)/(2 N_c)$ and $C_A = N_c$.
We introduce the perturbative expansion of $F_1$ and $F_2$ as
\begin{eqnarray}
  F &=& F^{(0)} + \left(\frac{\alpha_s(\mu)}{\pi}\right) F^{(1)}
        + \left(\frac{\alpha_s(\mu)}{\pi}\right)^2 F^{(2)} 
        + \cdots
  \,,
  \label{eq::F}
\end{eqnarray}
and decompose them into ``triangle'' and ``box'' form factors
\begin{eqnarray}
  F_1^{(k)} &=& \frac{3 m_H^2}{s-m_H^2} F^{(k)}_{\rm tri}+F^{(k)}_{\rm box1}
                \,, \nonumber\\
  F_2^{(k)} &=& F^{(k)}_{\rm box2}\,.
                \label{eq::F_12}
\end{eqnarray}
Note that for $t=0$ we have $F_2=0$, which we have explicitly checked in our
calculation. In this work we concentrate on $F^{(2)}_{\rm box1}$ since the
three-loop triangle form factor has already been computed in
Refs.~\cite{Davies:2019nhm,Davies:2019roy,Harlander:2019ioe,Czakon:2020vql}.

Leading-order predictions to $gg\to HH$ are known from
Refs.~\cite{Glover:1987nx,Plehn:1996wb} and NLO corrections based on a purely
numerical approach have been obtained in
Ref.~\cite{Borowka:2016ehy,Borowka:2016ypz,Baglio:2018lrj}.  In
Refs.~\cite{Bellafronte:2022jmo,Davies:2023vmj} it has been shown that the
combination of deep high-energy
expansions~\cite{Davies:2018ood,Davies:2018qvx,Davies:2023vmj} and results
from the expansion for small transverse
momentum~\cite{Bonciani:2018omm,Davies:2023vmj} lead to precise results for
the two-loop form factors which can be evaluated numerically in a fast and
flexible way.

Predictions at NNLO and N$^3$LO are mainly restricted to the large top quark
mass limit (see, e.g. Refs.~\cite{deFlorian:2013jea,Grigo:2014jma,Grazzini:2018bsd,Chen:2019lzz,Chen:2019fhs}) with the exception of Ref.~\cite{Davies:2023obx} where the
three-loop light-fermion contributions have been computed for $t=0$ and
$m_H=0$ but taking into account the full dependence on $s/m_t^2$.  Recently
also the leading logarithmic high-energy behaviour of the 
form factors has been studied in Ref.~\cite{Jaskiewicz:2024xkd}.

The remainder of the paper is organized as follows: In the next section we
present  details of our calculations and discuss the various
challenges.  Afterwards, in Section~\ref{sec::UVIR} we briefly discuss the
ultraviolet and infrared behaviour of the three-loop form factor. We also
discuss the dependence of our form factor on the renormalization scales
for the strong coupling constant and the top quark mass.
We present the results for the form factors in Section~\ref{sec::results}
and discuss in particular the reduction of the scheme dependence.
We conclude in Section~\ref{sec::concl} and give a brief outlook.


\section{Technical details}

The calculation follows the same strategy as has been outlined in Ref.~\cite{Davies:2023obx}
for the light fermion contributions. 
However, in order to perform the calculation for the large-$N_c$ contributions 
several improvements had to be implemented.

We generate the diagrams required for our calculation with \texttt{qgraf}~\cite{Nogueira:1991ex},
and then use \texttt{tapir}~\cite{Gerlach:2022qnc} and \texttt{exp}~\cite{Harlander:1998cmq,Seidensticker:1999bb} 
to map the diagrams onto topologies in full kinematics and convert the output to 
\texttt{FORM}~\cite{Ruijl:2017dtg} notation. 
The diagrams are then computed with the in-house
``{\tt calc}'' setup, to produce an amplitude in terms of scalar Feynman integrals
in a highly automated way.

Next we perform the expansions for $p_T \to 0$. 
The first step is to expand in the external Higgs mass $m_H$.
In this work we only consider the leading term in the $p_T \to 0$ expansion, 
thus we can simply set $m_H=0$ and do not have to perform a 
non-trivial expansion as has been done, e.g., in Refs.~\cite{Davies:2018qvx,Davies:2020lpf,Davies:2025wke}.

The second step is to expand the amplitude in the forward-scattering limit (i.e.~$t \to 0$).
Although we are also currently only interested in the leading term in this limit 
we have to perform a non-trivial expansion due to factors of $1/t$ which 
are present in the Lorentz projectors of the form factors. 
We implement this expansion in \texttt{FORM} by introducing the vector 
$q_{\delta} = q_1 + q_3$ and expand around $q_{\delta}=0$.
We have that $q_{\delta}^2=t$ and use a tensor reduction procedure to treat numerators of 
$q_{\delta}$ contracted with loop momenta.
After expanding the denominators they can become linearly dependent;
we have to perform a partial fraction decomposition before the subsequent integration-by-parts
(IBP) reduction to master integrals.
We produce the necessary identities in two independent ways, with \texttt{tapir} and \texttt{Limit}~\cite{Herren:2020ccq}, by specifying the kinematics $q_3=-q_1$.
The amplitude is now given in terms of scalar loop integrals in forward kinematics.

The procedure outlined above results in 522 integral families which are not all
independent. Our final simplification of the amplitude before IBP reduction is to find mappings
between these families using \texttt{feynson}~\cite{Maheria:2022dsq,feynson}, which results in an amplitude in
terms of 203 independent integral families. We note here that the excellent performance
of \texttt{feynson} was crucial to perform this step in a reasonable time.

At this point our amplitude is written in terms of around 2.6 million integrals belonging to the
203 independent families, which depend on the variables $d$ and $s/m_t^2$.
We perform the IBP reduction for each family independently with
\texttt{Kira}~\cite{Maierhofer:2017gsa,Klappert:2020nbg}.
The reduction of some non-planar integral families with rank-$5$ numerators is very computationally expensive 
and had to be performed on machines with up to $4$TB of RAM.
After reducing each family individually we obtain reduction tables in terms of over 33 thousand
master integrals, far too  many to compute. The next procedure is to minimize the number of master
integrals between the families to find a linearly independent basis. Due to resource constraints we
were not able to simply reduce the master integrals between all families using \texttt{Kira}. Therefore we
use the following procedure.

The first step is to use \texttt{FIRE}'s \texttt{FindRules} routine to find 1:1 maps between the 33
thousand integrals, which yields a basis of 4313 integrals. We then apply (a parallelized version of)
\texttt{FindRules} to the full
list of 2.6 million integrals of the amplitude to yield 1.3 million equivalent pairs. Applying the
IBP tables to these pairs and comparing the left- and right-hand sides yields 820 thousand non-trivial
relations involving 4029 of the 4313 basis integrals. Using \texttt{Kira}'s \texttt{user_defined_system}
to find reduction relations for the redundant integrals yields a basis of 1647 master integrals.
However, the differential system w.r.t.~$s/m_t^2$ for this basis contains some unpleasant features,
suggesting that this basis still contains linearly-dependent integrals; for example, it contains coupled
systems between integrals from different sectors. To eliminate additional integrals from the basis we
perform ``test reductions'' of a restricted integral list with \texttt{FIRE 6.5} \cite{Smirnov:2023yhb}
(where we find excellent performance using the \texttt{FUEL} \cite{Mokrov:2023vva} interface to the
\texttt{FLINT}~\cite{flint} library for rational-polynomial simplification),
and repeating the above
steps using these reduction tables yields 35 thousand, 1817 and finally 1561 master integrals for which
the differential system has a much improved structure (though the basis is likely still not fully minimal).

The calculation of this final set of 1561 master integrals poses a significant computational 
challenge, which we postpone for later study. 
In order to obtain first phenomenological relevant results we restrict ourselves 
to the leading-colour approximation, i.e.~$C_A \to N_c$, $C_F \to N_c/2$. 
This reduces the set of necessary master integrals to 783, which we were able 
to tackle with the ``expand and match'' method~\cite{Fael:2021kyg,Fael:2021xdp,Fael:2022miw}.
We note that the leading-colour limit does not eliminate non-planar topologies from 
our amplitude, as e.g.~shown in Fig.~\ref{fig::gghh_FD2}(c).
Conceptionally the method is also able to address the calculation of the 
full set of master integrals, however, computational bottlenecks have to be 
overcome first. 
In Sec.~\ref{sec::results} we comment on the quality of the large-$N_c$ approximation 
in the large-$m_t$ limit.

We compute the master integrals with the help of their differential equations with 
respect to $x = s/m_t^2$. 
First, we construct analytic results in the large-$m_t$ ($x \to 0$) limit.
To achieve this we first insert an ansatz for the master integrals expanded around $x \to 0$ 
into the differential equation. This leads to a large system of linear equations for 
the expansion coefficients which we solve in terms of a small number of boundary coefficients. 
In order to fix these coefficients we compute the first few terms in the large-$m_t$ expansion explicitly. 
The calculation is facilitated by \texttt{exp}, which automates the 
asymptotic expansion in the limit $m_t^2 \gg s$. 
The different regions lead to three-loop vacuum integrals, as well as products of one- 
and two-loop vacuum integrals with two- and one-loop massless $s$-channel vertex integrals, respectively,
which are all well known in the literature.
We only need about half of the computed expansion terms to fix the boundary conditions 
of our symbolic expansion; the rest are used for welcome consistency checks of our calculation.
After inserting the master integrals into the amplitude we reproduce the results of
Ref.~\cite{Davies:2019djw} after specifying to the large-$N_c$ limit.

In a subsequent step we use the ``expand and match'' method to transport the 
results valid around $s/m_t^2 \to 0$, and numerical evaluations at other points,
to  nearby values~\cite{Fael:2021kyg,Fael:2021xdp,Fael:2022miw}.
Compared to the calculation in Ref.~\cite{Davies:2023obx} this step is much more complex;
the system of equations is much larger, and exhibits various unphysical singularities in $x$ which
limit the radii of convergence of the expansions.
We therefore had to perform expansions around a much larger number of points:
\begin{eqnarray}
    \frac{s}{m_t^2} &=& x \in \left\{ 
        0,1,2,3,\frac{7}{2},4,\frac{9}{2},5,
        6,7,8,10,12,14,15,16,18,22,27,35,45
    \right\}~.
\end{eqnarray}
We use \texttt{AMFlow}~\cite{Liu:2017jxz,Liu:2021wks,Liu:2022mfb,Liu:2022chg} 
at the values $x = \{ 1,5,10,27 \}$ with 40 digits accuracy and match them directly to symbolic 
expansions around these points. 
Expansions around other points are obtained by  the ``expand and match'' method.
We verify that the expansions obtained after crossing the physical cuts at $x=4$ (the two-particle threshold) and $x=16$
(the four-particle threshold),\footnote{We observe that the final result for the form factor in the large-$N_c$ approximation has no cut at $x=16$. However, we have used a general ansatz in the ``expand and match'' approach which allows for square roots in $(s/m_t^2-16)$. All half-integer powers of $(s/m_t^2-16)$ cancel in the matching process.}
numerically agree with at least 
10 significant digits with the ones obtained from expansions obtained by 
matching with \texttt{AMFlow} runs above the cuts.
This provides a strong check on the analytic continuation over these 
singular points of the differential equation and on the quality of the 
semi-numerical solutions over the whole range of $x$.

The \texttt{AMFlow} runs have been facilitated by the use of \texttt{Symbolica}~\cite{ruijl_2025_15040848}
instead of \texttt{Fermat}~\cite{fermat} as the back-end of a modified version of \texttt{Kira}
to perform the simplification of rational polynomials.
This modification reduces the run time for a numerical evaluation of the most complicated integral families 
from over one month to less than a week, with similar computing 
resources.
Using this approach we obtain smooth transitions between the different expansions with an agreement of ten or more
digits at the matching points.
The results provide us an accuracy of about 10 significant digits for the finite part of the form factor for $\sqrt{s}\lesssim 1000$~GeV. If
required an extension to higher energies is possible.


\section{\label{sec::UVIR}Form factors in the $\overline{\rm MS}$ and pole scheme}

{\bf Ultraviolet renormalization}


We first renormalize the top quark mass in the pole ($M_t$) or
$\overline{\rm MS}$ ($\overline{m}_t$) scheme and the strong coupling constant
in the $\overline{\rm MS}$ scheme with six active flavours. This requires that
we renormalize the gluon wave function for which we also use the
pole scheme, i.e. the gluon two-point function with external momentum squared evaluated to zero.  All counterterms are needed to two-loop order.  For
completeness we provide the explicit expressions (see, e.g., Refs.~\cite{Gray:1990yh,Chetyrkin:2004mf,Gerlach:2018hen}): 
\newcommand{\lmmOS}{l_{\rm OS}}
\newcommand{\lmm}{l_{\rm \overline{\rm MS}}}
\newcommand{\ep}{\epsilon}
\newcommand{\api}{\frac{\alpha_s^{(6)}(\mu)}{\pi}}
\begin{eqnarray}
  Z_{m_t}^{\rm OS} &=& 
  1 
  +\left(\frac{\alpha_s^{(6)}(\mu)}{\pi}\right) C_F \biggl\{
        -\frac{3}{4 \epsilon }
        -\frac{3}{4} \lmmOS
        -1
        + \epsilon  \biggl[-\frac{3 \lmmOS^2}{8}-\lmmOS-2-\frac{\pi ^2}{16}\biggr] 
        \nonumber \\ &&
        + \epsilon ^2 \biggl[
                -\frac{\lmmOS^3}{8}
                -\frac{\lmmOS^2}{2}
                -\frac{1}{16} \lmmOS 
                \bigg(
                        32
                        +\pi ^2
                \bigg)
                -4             
                -\frac{\pi ^2}{12}
                +\frac{\zeta_3}{4}
        \biggr]
    \biggr\}
    \nonumber \\ &&
    + \left( \frac{\alpha_s^{(6)}(\mu)}{\pi} \right)^2 
    \biggl\{
        \frac{1}{\epsilon ^2} \biggl[
                \frac{11 C_A C_F}{32}
                +\frac{9 C_F^2}{32}
                -\frac{1}{8} C_F ( n_h + n_l ) T_F
        \biggr]
        +\frac{1}{\epsilon } \biggl[
                -\frac{97}{192} C_A C_F
                \nonumber \\ &&
                +\frac{9}{64} (5+4 \lmmOS) C_F^2
                +\frac{5}{48} C_F (n_h + n_l) T_F
        \biggr]
        +C_F T_F n_l \biggl[
                 \frac{\lmmOS^2}{8}
                +\frac{13 \lmmOS}{24}
                +\frac{71}{96}
                +\frac{\pi ^2}{12}
        \biggr]
        \nonumber \\ &&
        +C_F T_F n_h \biggl[
                 \frac{\lmmOS^2 }{8} 
                +\frac{13\lmmOS}{24} 
                +\frac{143}{96}
                -\frac{1}{6} \pi ^2 
        \biggr]
        +C_F^2 \biggl[
                 \frac{9 \lmmOS^2}{16}
                 +\frac{45 \lmmOS}{32}
                +\frac{199}{128}
                -\frac{17 \pi ^2}{64}
                \nonumber \\ &&
                +\frac{1}{2} \pi ^2 \log (2)
                -\frac{3 \zeta_3}{4}
        \biggr]
        +C_A C_F \biggl[
                -\frac{11 \lmmOS^2}{32}
                -\frac{185 \lmmOS}{96}
                -\frac{1111}{384}                
                +\frac{\pi ^2}{12}
                \nonumber \\ &&
                -\frac{1}{4} \pi ^2 \log (2)
                +\frac{3 \zeta_3}{8}
        \biggr]
    \biggr\}
                       \,,\nonumber\\
  Z_{m_t}^{\overline{\rm MS}} &=& 
  1
  -\frac{3 C_F}{4 \epsilon } \left(\frac{\alpha_s^{(6)}(\mu)}{\pi}\right)
  + \left(\frac{\alpha_s^{(6)}(\mu)}{\pi}\right)^2 
  \biggl\{
        \frac{1}{\epsilon ^2} \biggl[
                \frac{11 C_A C_F}{32}
                +\frac{9 C_F^2}{32}
                -\frac{1}{8} C_F (n_h+n_l) T_F
        \biggr]
        \nonumber \\ &&
        +\frac{1}{\epsilon } \biggl[
                -\frac{97}{192} C_A C_F
                -\frac{3 C_F^2}{64}
                +\frac{5}{48} C_F (n_h+n_l) T_F
        \biggr]
    \biggr\}
                       \,,\nonumber\\
  Z_{\alpha_s}^{\overline{\rm MS}} &=& 1
    + \left(\frac{\alpha_s^{(6)}(\mu)}{\pi}\right)  \frac{1}{\epsilon}
    \biggl\{
        -\frac{11 C_A}{12}
        + \frac{T_F}{3} (n_h+n_l) 
    \biggr\}
    + \left(\frac{\alpha_s^{(6)}(\mu)}{\pi}\right)^2 \biggl\{
         \frac{1}{\epsilon ^2} \biggl[
                \frac{T_F^2 (n_h + n_l)^2}{9}
                \nonumber \\ &&
                + \frac{121 C_A^2}{144}
                - \frac{11}{18} C_A T_F (n_h+n_l)
        \biggr]
        +\frac{1}{\epsilon } \biggl[
                T_F (n_h + n_l) \biggl(
                        \frac{5 C_A }{24}
                        +\frac{C_F }{8}
                \biggr)
                -\frac{17 C_A^2}{48}
        \biggr]
    \biggr\}
                 \,,\nonumber\\
  Z_3^{\rm OS} &=& 1 
  + \left(\api\right) T_F {n_h}
  \biggl\{
    - \frac{1}{3\ep} - \frac{\lmm}{3}
    - \ep \left[\frac{\lmm^2}{6} + \frac{\pi^2}{36} \right]
    + \ep^2 \left[-\frac{\lmm^3}{18} - \frac{\pi^2\lmm}{36} + \frac{\zeta_{3}}{9}\right]
    \biggr\}
  \nonumber\\&&
  + \left(\api\right)^2 T_F { n_h}
  \Biggl\{
    \frac{1}{\ep^2}\left[ \frac{35 C_F }{144} -  \frac{T_F n_l}{9} \right]
    + \frac{1}{\ep} \biggl[
          C_A \frac{{ 13} \lmm}{72}
        + T_F { (n_h -n_l) }\frac{\lmm}{9}
        \nonumber\\&&
        -  \frac{5 C_A}{32} 
        -  \frac{C_F}{8} 
        \biggr]
        + T_F { n_h}
        \bigg( 
              \frac{ \lmm^2 }{6}
            + \frac{\pi^2}{108}
        \bigg)
        - T_F n_l 
        \bigg( 
              \frac{ \lmm^2 }{18}
            +  \frac{\pi^2}{108}
        \bigg)
        + C_A 
        \bigg( 
            \frac{\lmm^2 }{36}
            -\frac{5\lmm }{16}
            \nonumber\\&&
            +\frac{13}{192}
            +\frac{13 \pi^2}{864}
        \bigg)
        + C_F 
        \bigg(
              \frac{ \lmm}{4}
            - \frac{13}{48}
        \bigg)
       \Biggr\}
          \,,
                   \label{eq::Z}
\end{eqnarray}
with $\lmm=\log(\mu^2/\overline{m}_t^2)$
and $\lmmOS = \log(\mu^2/M_t^2)$.
The two-loop $gg\to HH$ amplitude develops $1/\epsilon^2$ poles which is why
we need the one-loop expressions of $Z_{m_t}^{\rm OS}$ and $Z_3^{\rm OS}$ to order
$\epsilon^2$.  The one-loop $gg\to HH$ amplitude is finite and thus constant
terms in $\epsilon$ are sufficient at order $\alpha_s^2$.

We next decouple the contribution of the top quark from the running
of $\alpha_s$ and express our amplitude in terms of $\alpha_s^{(5)}(\mu_s)$.
The corresponding decoupling constant defined via
\begin{eqnarray}
  \alpha_s^{(5)}(\mu_s) &=& \alpha_s^{(6)}(\mu_s)\: \zeta_{\alpha_s}
                            \,,
\end{eqnarray}
is given by
\begin{eqnarray}
  \zeta_{\alpha_s}^{\rm OS} &=& 1
  + \left(\frac{\alpha_s^{(5)}(\mu)}{\pi}\right)
  T_F { n_h} \left[ 
    \frac{\lmmOS}{3} 
    + \ep\left( \frac{\lmmOS^2}{6} + \frac{\pi^2}{36} \right)
    + \ep^2\left(\frac{\lmmOS^3}{18} + \frac{\pi^2\lmmOS}{36} - \frac{\zeta_{3}}{9}\right)
   \right]
  \nonumber\\&&
  + \left(\frac{\alpha_s^{(5)}(\mu)}{\pi}\right)^2 { T_F n_h}
  \left[
  C_F \left(\frac{15}{16} + \frac{\lmmOS}{4} \right)
  + C_A \left(-\frac{2}{9} + \frac{5\lmmOS}{12} \right)
  + T_F n_h \frac{\lmmOS^2}{9}
  \right] 
                       \,.
\end{eqnarray}
The one-loop expression again needs to include $\epsilon^2$ terms.  The
quantity $\zeta_{\alpha_s}^{\overline{\rm MS}}$ is obtained with the help of
the renormalization constants in Eq.~(\ref{eq::Z}).


{\bf Infrared subtraction}


For the subtraction of the infrared poles we adapt the same procedure as
in Ref.~\cite{Davies:2019djw} which is based on Ref.~\cite{Catani:1998bh}, see
also Refs.~\cite{deFlorian:2012za,Grigo:2014jma}.

Finite form factors at NLO and NNLO are obtained via the following 
subtraction procedure
\begin{eqnarray}
  F^{(1),\rm fin} &=& F^{(1)} - \frac{1}{2} I^{(1)}_g F^{(0)}\,,\nonumber\\
  F^{(2),\rm fin} &=& F^{(2)} - \frac{1}{2} I^{(1)}_g F^{(1)}
  - \frac{1}{4} I^{(2)}_g F^{(0)}\,,
  \label{eq::FF_IR}
\end{eqnarray}
where the quantities on the right-hand side are ultraviolet-renormalized and
$I^{(1)}_g$ and $I^{(2)}_g$ are given by~\cite{Catani:1998bh,deFlorian:2012za}
\begin{eqnarray}
	I^{(1)}_g &=&
		{} - \left(\frac{\mu^2}{-s-i\delta}\right)^\epsilon
		\frac{e^{\epsilon\gamma_E}}{\Gamma(1-\epsilon)}
		\frac{1}{\epsilon^2}
		\Big[
			C_A + 2\epsilon\beta_0
		\Big]\,,\\
	I^{(2)}_g &=&
		{} - \left(\frac{\mu^2}{-s-i\delta}\right)^{2\epsilon}
		\left(\frac{e^{\epsilon\gamma_E}}{\Gamma(1-\epsilon)}\right)^2
		\frac{1}{\epsilon^4}
		\Big[
			\frac{1}{2}(C_A+2\epsilon\beta_0)^2
		\Big]\nonumber\\
		&& {} + \left(\frac{\mu^2}{-s-i\delta}\right)^{\epsilon}
		\frac{e^{\epsilon\gamma_E}}{\Gamma(1-\epsilon)}
		\frac{1}{\epsilon^3}
		\Big[
			2(C_A+2\epsilon\beta_0)\beta_0
		\Big]\nonumber\\
		&& {} - \left(\frac{\mu^2}{-s-i\delta}\right)^{2\epsilon}
		\frac{e^{\epsilon\gamma_E}}{\Gamma(1-\epsilon)}
		\bigg\{
			\frac{1}{\epsilon^3}
			\Big[
				\frac{1}{2}(C_A+4\epsilon\beta_0)\beta_0
			\Big]\nonumber\\
			&& \hspace{2cm}
			- \frac{1}{\epsilon^2}
			\Big[
				\frac{(3\pi^2-67)C_A + 10n_l}{72}(C_A+4\epsilon\beta_0)
			\Big]
			- \frac{1}{\epsilon}
			\Big[
				\frac{1}{2}H_g
			\Big]
		\bigg\}\,,
                           \label{eq::I1I2}
\end{eqnarray}
with
\begin{eqnarray}
\label{eq::b0Hgdef}
  \beta_0 &=& \frac{1}{4}\left(\frac{11}{3}C_A - \frac{4}{3} T n_l \right)\,,
              \nonumber\\
  H_g &=&
  C_A^2\left(\frac{\zeta_3}{2}+\frac{5}{12}+\frac{11\pi^2}{144}\right)
  + C_A n_l \left( \frac{29}{27} + \frac{\pi^2}{72} \right)
  + \frac{1}{2} C_F n_l + \frac{5}{27} n_l^2
  \,.
\end{eqnarray}

We compute all counterterm contributions and infrared subtraction
terms for general colour factors. Afterwards we select the leading-$N_c$ terms for the contributions with one closed top quark loop.
Furthermore we add the light-fermion contributions (``$n_l$'')
computed in Ref.~\cite{Davies:2023obx}. The renormalization and
infrared subtraction procedure also produces
$n_l^2$ terms, which we take into account. These (which are
numerically small even for $n_l=5$) cancel
against the convolutions with splitting functions which
are part of the real radiation contribution.


{\bf Renormalization scales}


Note that the subtraction terms in Eq.~(\ref{eq::I1I2}) are chosen in such a
way that the renormalization scale dependence of $F^{\rm fin}$ is governed by
$\alpha_s \equiv \alpha_s(\mu_s)$ in case the top quark mass is renormalized
on-shell.  If the top quark mass is renormalized in the $\overline{\rm MS}$
scheme we introduce in a first step $\overline{m}_t(\mu_s)$. Afterwards we use
the two-loop renormalization group equation for $\alpha_s$ to separate the
scales and express the form factors in terms of $\alpha_s(\mu_s)$ and
$\overline{m}_t(\mu_t)$.

Note that each time we discuss the renormalization scale dependence of the
form factors we actually have to take into account the factor $\alpha_s$ in
Eq.~(\ref{eq::X0}) and have to consider the combination
$\alpha_s(\mu_s) F(\mu_s,\mu_t)$. For this reason, in the next section
we show results for the quantity
\begin{eqnarray}
    \label{eq:Gdef}
  G(\mu_s,\mu_t) &=& \frac{\alpha_s(\mu_s)}{\alpha_s(M_t)} F^{\rm fin}(\mu_s,\mu_t)\,,
\end{eqnarray}
where $M_t$ is the top quark pole mass.



\section{\label{sec::results}Scheme and scale dependence of the form factor at NNLO}

We use the same input values as in Ref.~\cite{Davies:2023obx}
which are given by
\begin{eqnarray}
  M_t &=& 173.21~\mbox{GeV}\,,  \nonumber\\
  m_H &=& 125.10~\mbox{GeV}\,,  \nonumber\\
  \alpha_s^{(5)}(M_Z) &=& 0.118\,.
\end{eqnarray}
For the numerical evaluation of the form factors we need
$\alpha_s^{(5)}(\mu_s)$ and $\overline{m}_t(\mu_t)$.  For the latter we also
need $\alpha_s^{(6)}(\mu_t)$.  The corresponding numerical values are obtained
using {\tt RunDec}~\cite{Herren:2017osy} with five-loop running and four-loop
matching at $\mu=M_t$
for the transition from the $n_f=5$ to the $n_f=6$
flavour theory. For reference we provide the values for
$\mu_t=\mu_s=\overline{m}_t(\overline{m}_t)$:
\begin{eqnarray}
  \overline{m}_t(\overline{m}_t) &=& 163.39~\mbox{GeV}\,,  \nonumber\\
  \alpha_s^{(6)}(\overline{m}_t(\overline{m}_t)) &=& 0.108\,.
\end{eqnarray}


\begin{figure}[t]
  \begin{tabular}{cc}
    \includegraphics[width=0.45\textwidth]{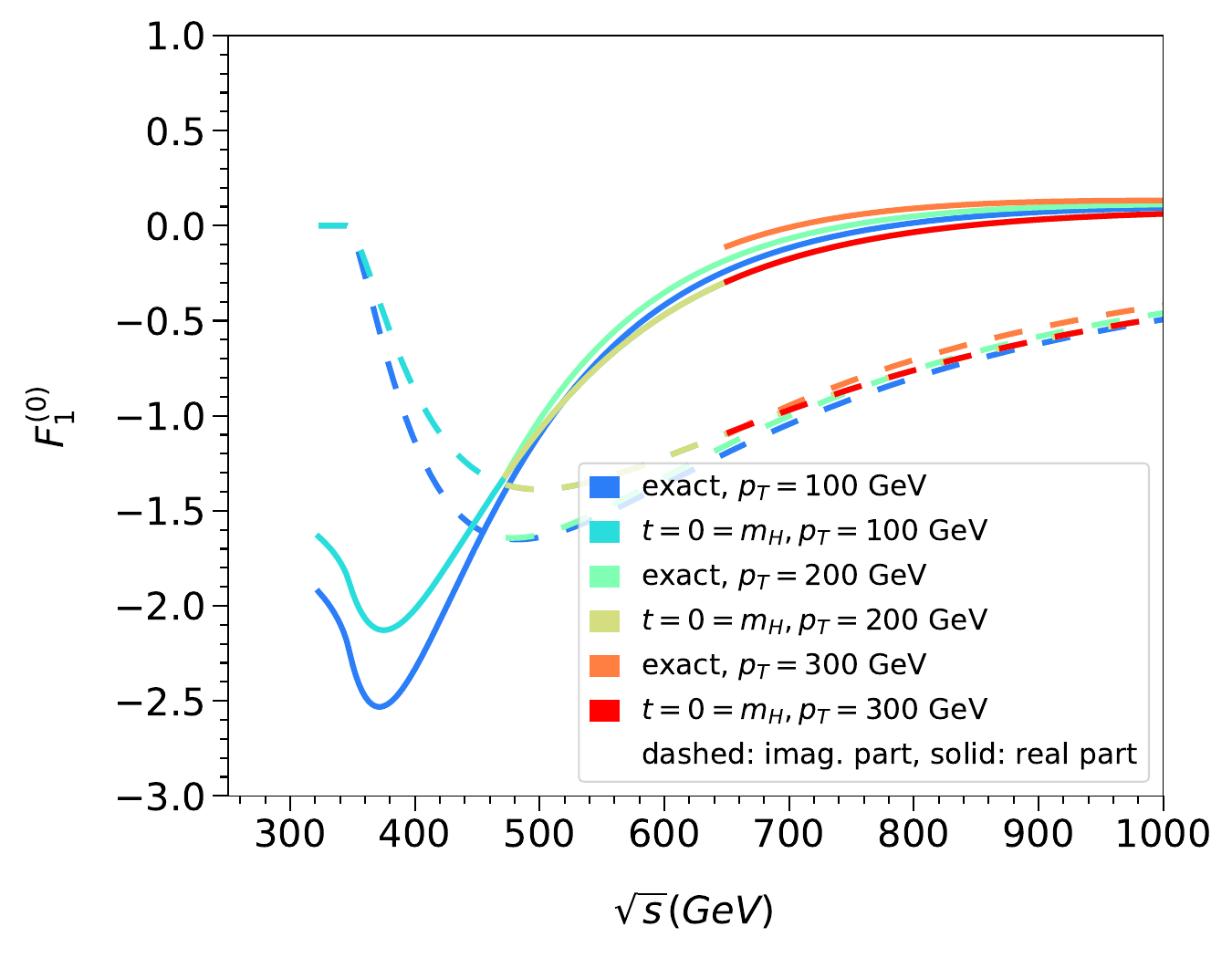} &
    \includegraphics[width=0.45\textwidth]{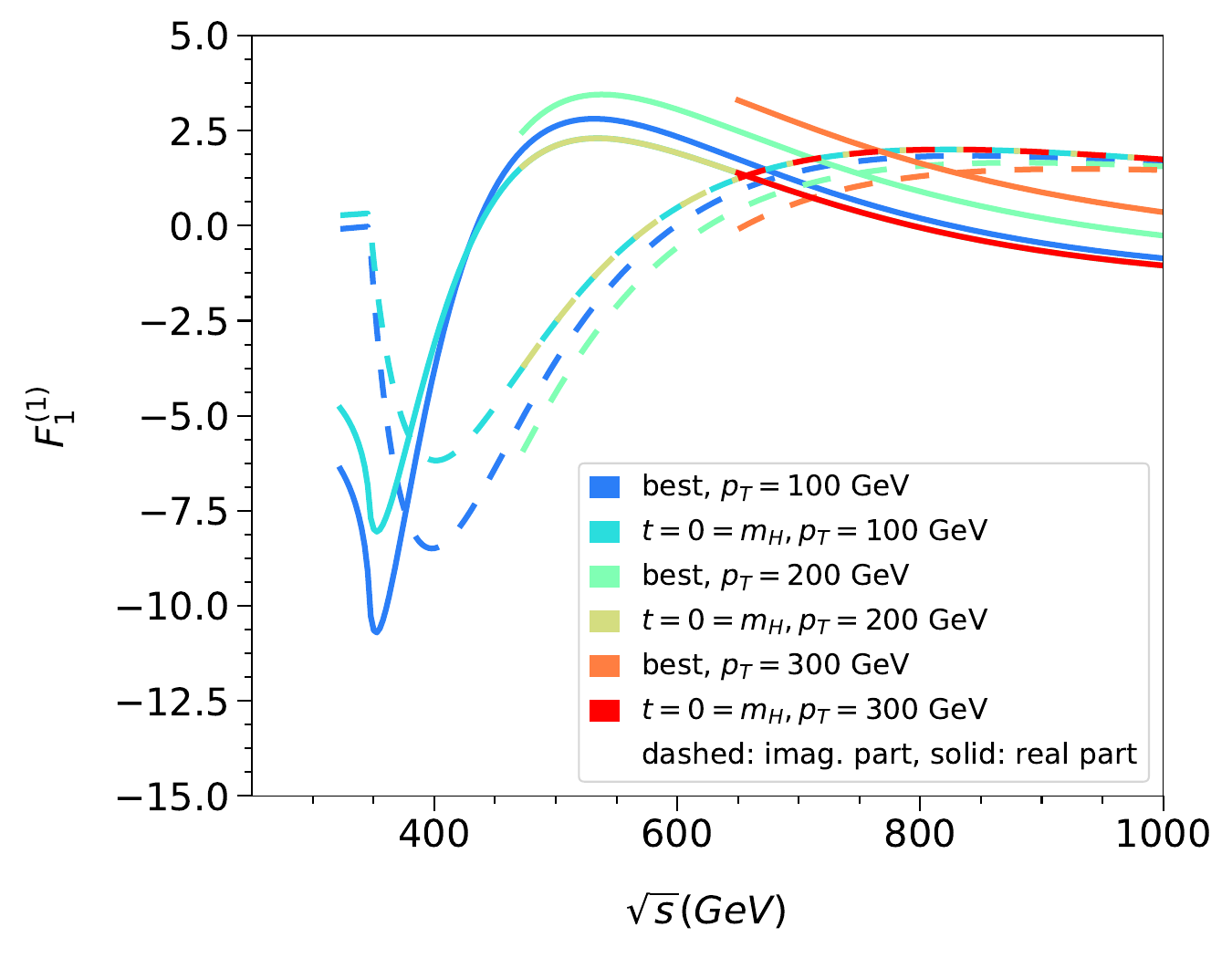}
  \end{tabular}
  \caption{\label{fig::F1_12l}One- and two-loop results for $F_1^{\rm fin}$. We show
    both the exact result and the approximation for $t=0$ and $m_H=0$ (which does not depend on $p_T$)
    for various values of $p_T$. For the two-loop curves we 
    choose $\mu^2=-s$ as in Ref.~\cite{Davies:2023obx}.
    The label ``best'' refers to the combination of the small-$t$ and high-energy
    expansions which have been shown to be equivalent to the exact result, see
Ref.~\cite{Davies:2023vmj}. 
  }
\end{figure}

We start in Fig.~\ref{fig::F1_12l} by showing the real and imaginary parts
of the one- and two-loop form factor $F_1^{\rm fin}$. The exact results are
compared to the approximation we apply at three loops ($t=0$ and
$m_H=0$). Note that this approximation is independent of $p_T$. For the exact
result we show curves for $p_T=100$~GeV, $p_T=200$~GeV and $p_T=300$~GeV.
Fig.~\ref{fig::F1_12l} is based on data from Ref.~\cite{Davies:2023obx}
where $\mu_s^2=-s$ has been chosen and the top quark mass has been
renormalized in the pole scheme.

In the range of $p_T$ which we consider we observe, both at one and two loops and both
for the real and imaginary parts, only a mild dependence on $p_T$.  It is
impressive that the forward approximation for
$m_H=0$ agrees well with the exact result.
Around the top--anti-top threshold we observe 
a deviation of only 20\%.  This suggests that the three-loop
approximation for $t=0$ and $m_H=0$ already provides phenomenologically
relevant results, since a large part of the 
total cross section is provided
for values of the transverse momentum around 100~GeV.

Next we discuss the quality of the large-$N_c$ approximation, which we can test
in the large-$m_t$ limit where results for all colour structures are
available~\cite{Davies:2019djw}. If we concentrate on the leading $N_c^2$ term of the contribution with one closed top quark loop and consider the leading term in the
$1/m_t$ expansion
and values for $\sqrt{s}$ and $p_T$ where the large-$m_t$
approximation is valid (i.e., $250$~GeV$\lesssim \sqrt{s} \lesssim 300$~GeV
and $p_T$ below about 50~GeV) we observe that the deviation between the full
results and the leading $N_c^2$ term in the real and imaginary parts of
$F_1^{\rm fin}$ is about 5\% and 30\%, respectively.
After including sub-leading terms in $1/m_t$
the agreement improves further
and reaches approximately 10\% for the 
imaginary part.
Thus, it can be expected that the major contribution is covered by
the large-$N_c$ term in the limit $p_T \to 0$ computed in this paper.

\begin{figure}[t]
  \begin{tabular}{cc}
    \includegraphics[width=0.45\textwidth]{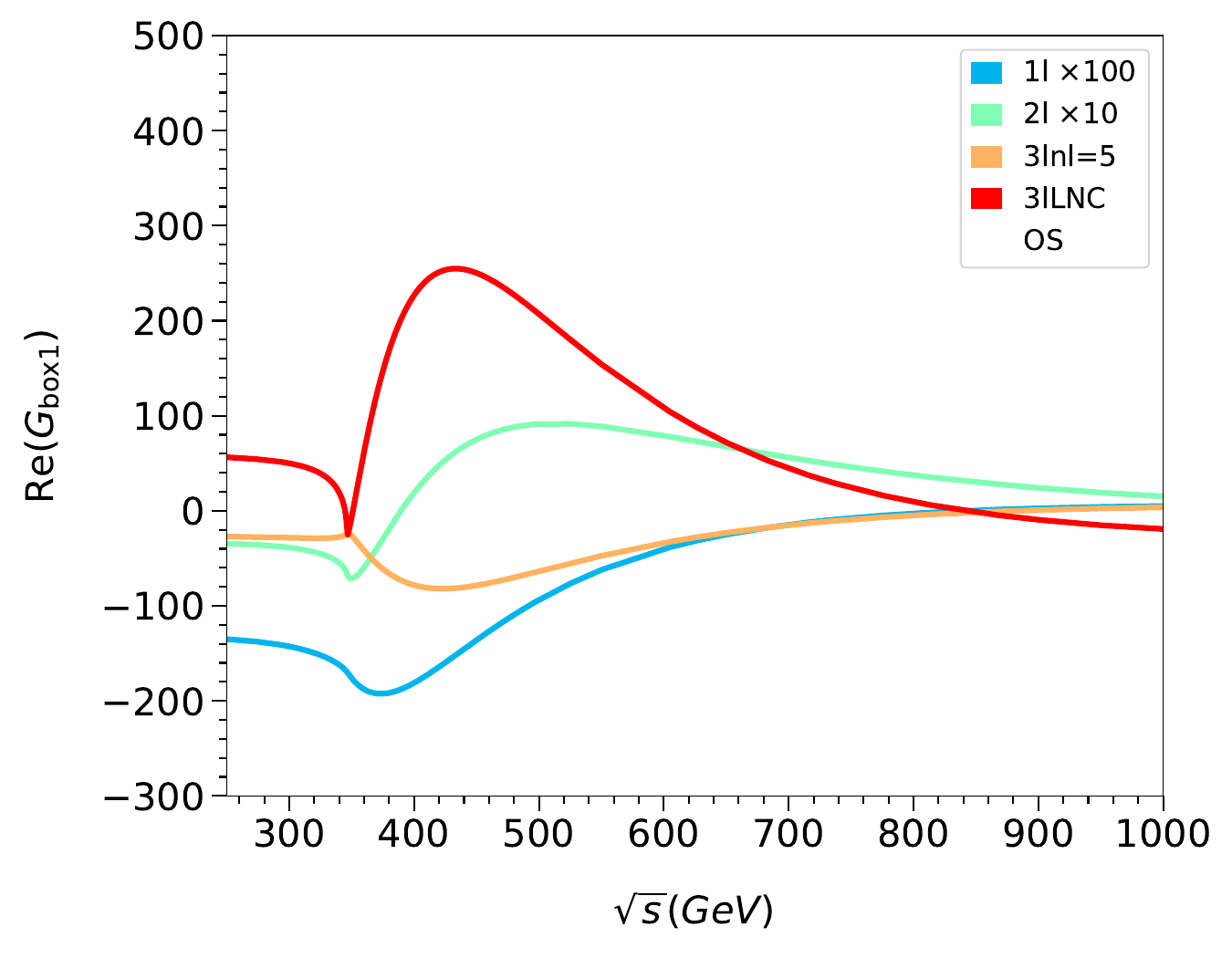} &
    \includegraphics[width=0.45\textwidth]{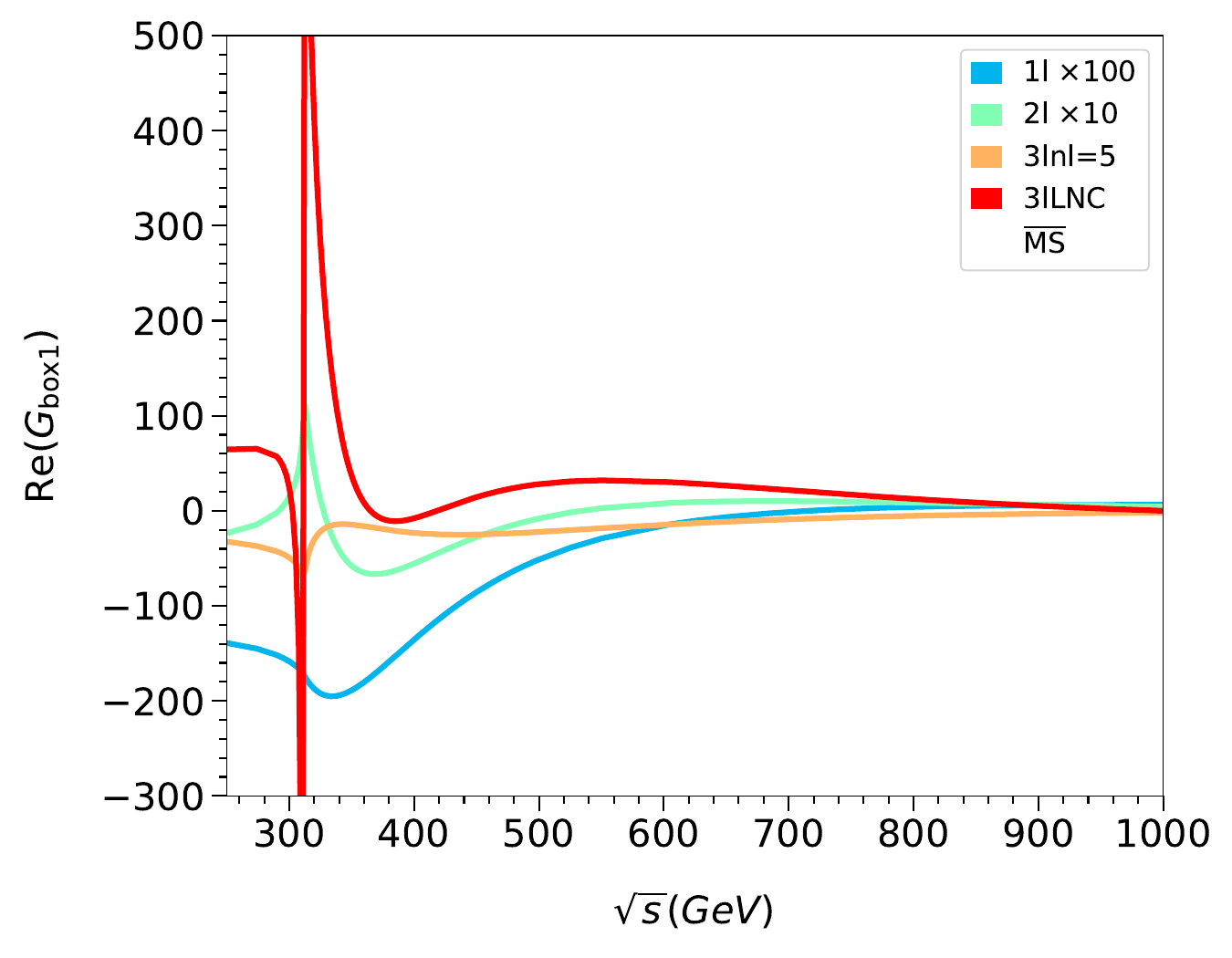}\\
    \includegraphics[width=0.45\textwidth]{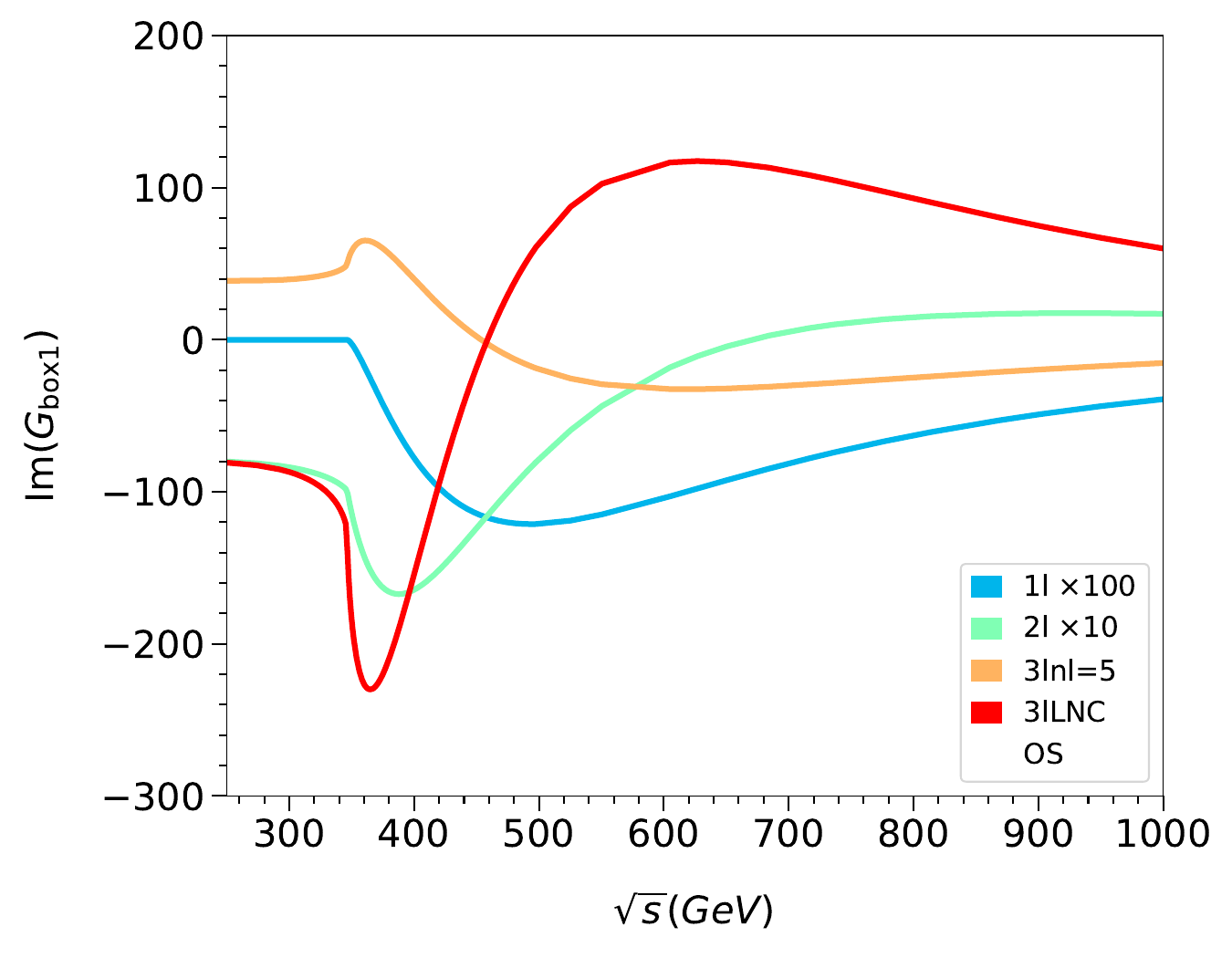} &
    \includegraphics[width=0.45\textwidth]{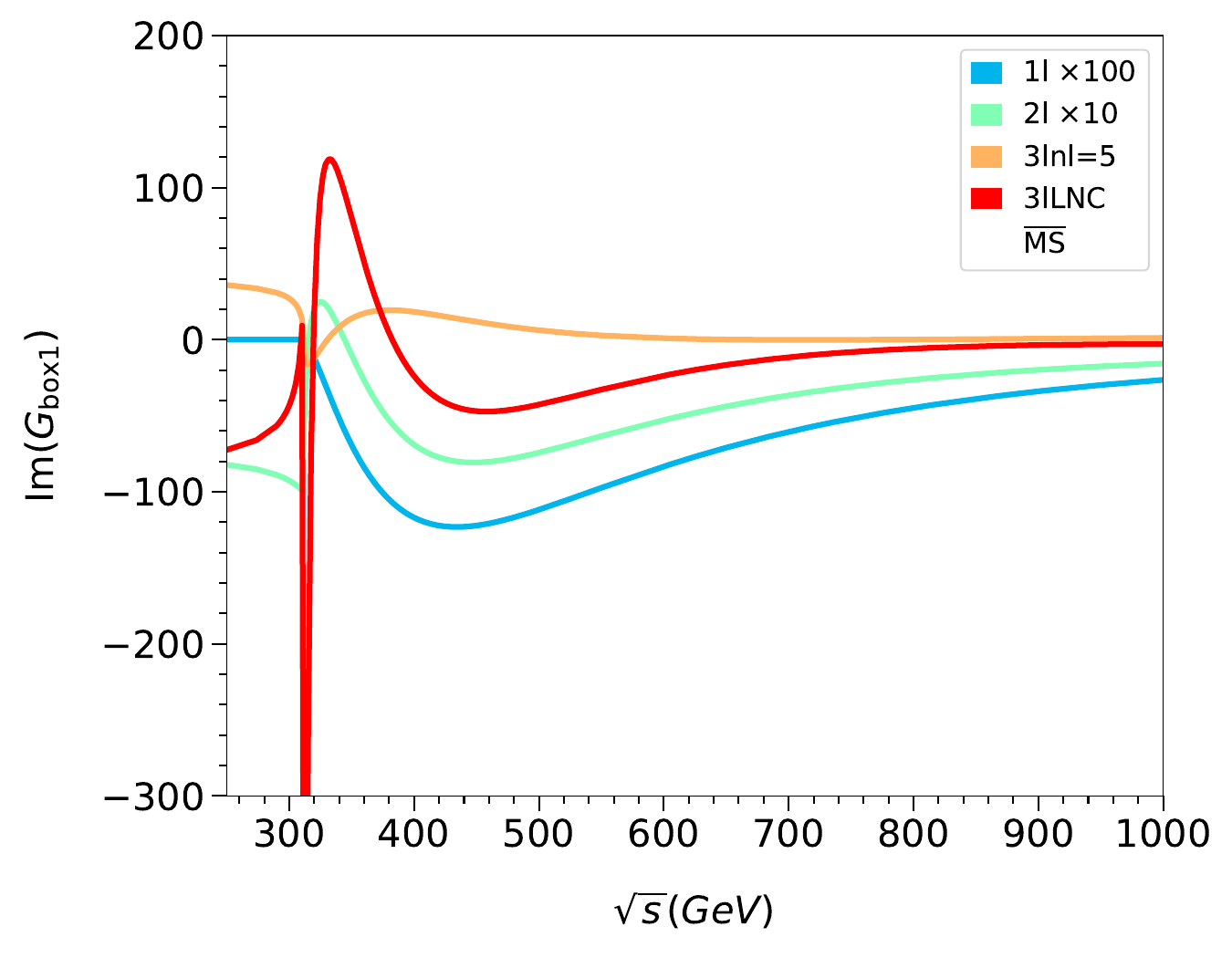}
  \end{tabular}
  \caption{\label{fig::F1_123} Real (top) and imaginary (bottom) parts of $G_{\rm box1}$
    for $t=0$ and $m_H=0$ at one, two and three loops. The panels on the left and right use the pole and $\overline{\rm MS}$ top quark mass, respectively.
    At three loops the
    light-fermion (for $n_l=5$) and large-$N_c$ contributions are shown
    separately. For the renormalization scales we
    have chosen $\mu_s^2=\mu_t^2=s$.}
\end{figure}

In Fig.~\ref{fig::F1_123} we
show the real and imaginary parts of the perturbative coefficients $G_{\rm box1}^{(0)}$,
$G_{\rm box1}^{(1)}$ and $G_{\rm box1}^{(2)}$  (as defined in Eq.~(\ref{eq:Gdef})) for the approximation $t=0$ and $m_H=0$.  For the
latter the light-fermion and large-$N_c$ results are shown separately. For the
renormalization scales we have chosen $\mu_s^2=\mu_t^2=s$. For clarity the
one- and two-loop results are multiplied by a factor 100 and 10, respectively.
In the left-hand column the pole scheme is used for the top quark mass and on the right-hand side we use
the $\overline{\rm MS}$ mass.

In the pole scheme we observe larger higher-order corrections.  Depending on
$\sqrt{s}$ the increase in the absolute value is in general more than an order of
magnitude.  This is a feature which is often observed in the pole scheme.  The
situation is different in the $\overline{\rm MS}$ scheme.  Here the higher-order
coefficients are much smaller which leads to a better convergence of
perturbation theory.

The $\overline{\rm MS}$ curves show a characteristic feature around
$\sqrt{s}=2\,\overline{m}_t\approx 320$~GeV which deserves an explanation.  The top quark mass
counterterm contributions from the one- and two-loop corrections are obtained
via derivatives with respect to $m_t$. At threshold, i.e.~for $s/m_t^2=4$, the derivatives are not analytic, which leads to numerically large contributions.
In the pole scheme the bare three-loop form factor
shows a similar behaviour with the opposite sign, such that at $s/m_t^2=4$ a
smooth behaviour is observed. On the other hand, if the top quark mass is
renormalized in the $\overline{\rm MS}$ scheme there is only a partial
cancellation which leads to the dip-peak structure as observed in
Fig.~\ref{fig::F1_123}. 
Note that a similar behaviour is also observed in hadronic quantities where often bins in
the Higgs boson pair invariant mass, $M_{HH}$, are used. Usually the bin including the top pair threshold shows large deviations between the $\overline{\rm MS}$ and pole
scheme, see, e.g., Fig.~2 of Ref.~\cite{Bagnaschi:2023rbx}.
In fact, close to threshold it is not
recommended to use the $\overline{\rm MS}$ definition for the top quark mass
so for practical purposes the non-physical behaviour of the form factor is
not a problem.



\begin{figure}[t]
  \begin{tabular}{cc}
    \includegraphics[width=0.45\textwidth]{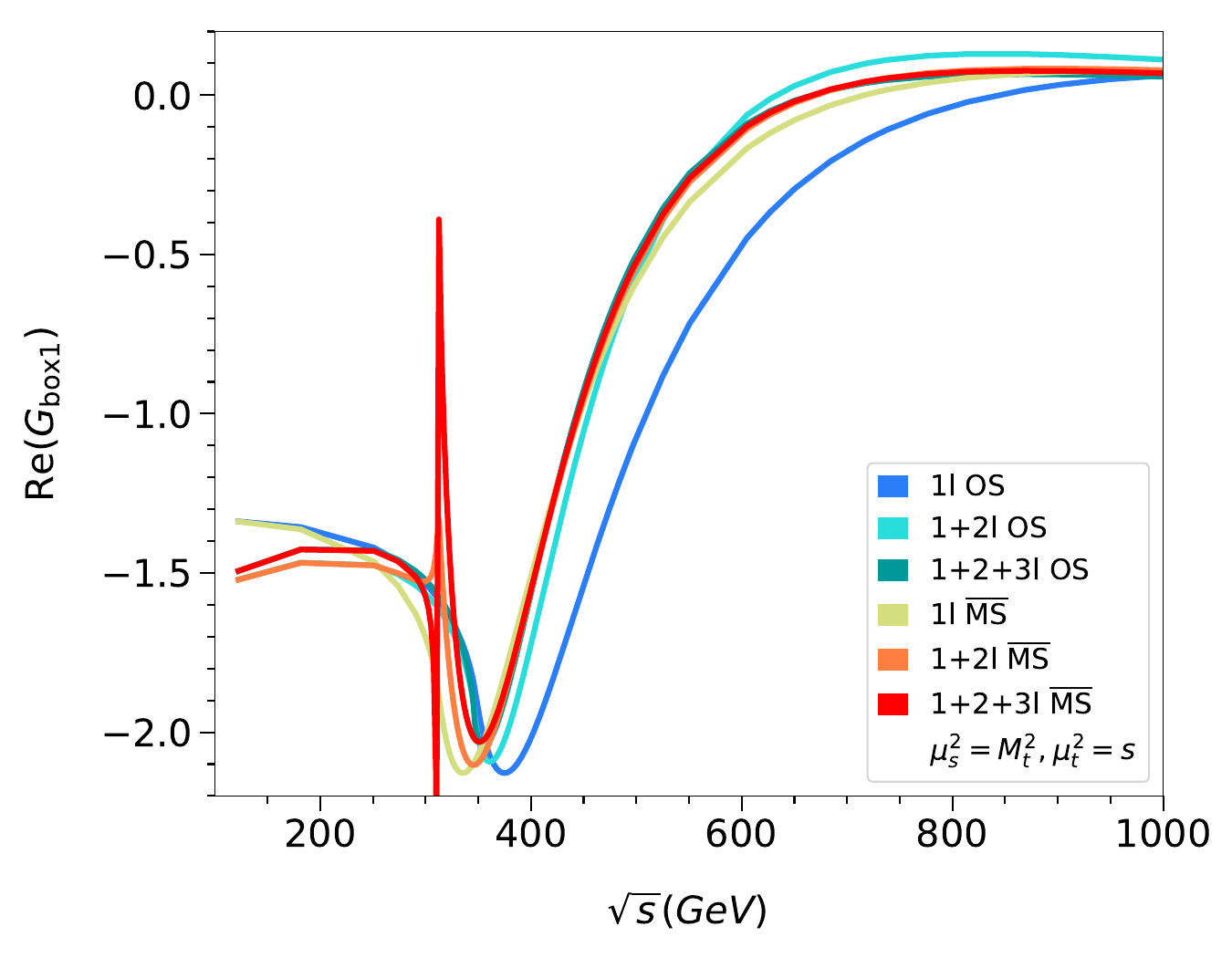} &
    \includegraphics[width=0.45\textwidth]{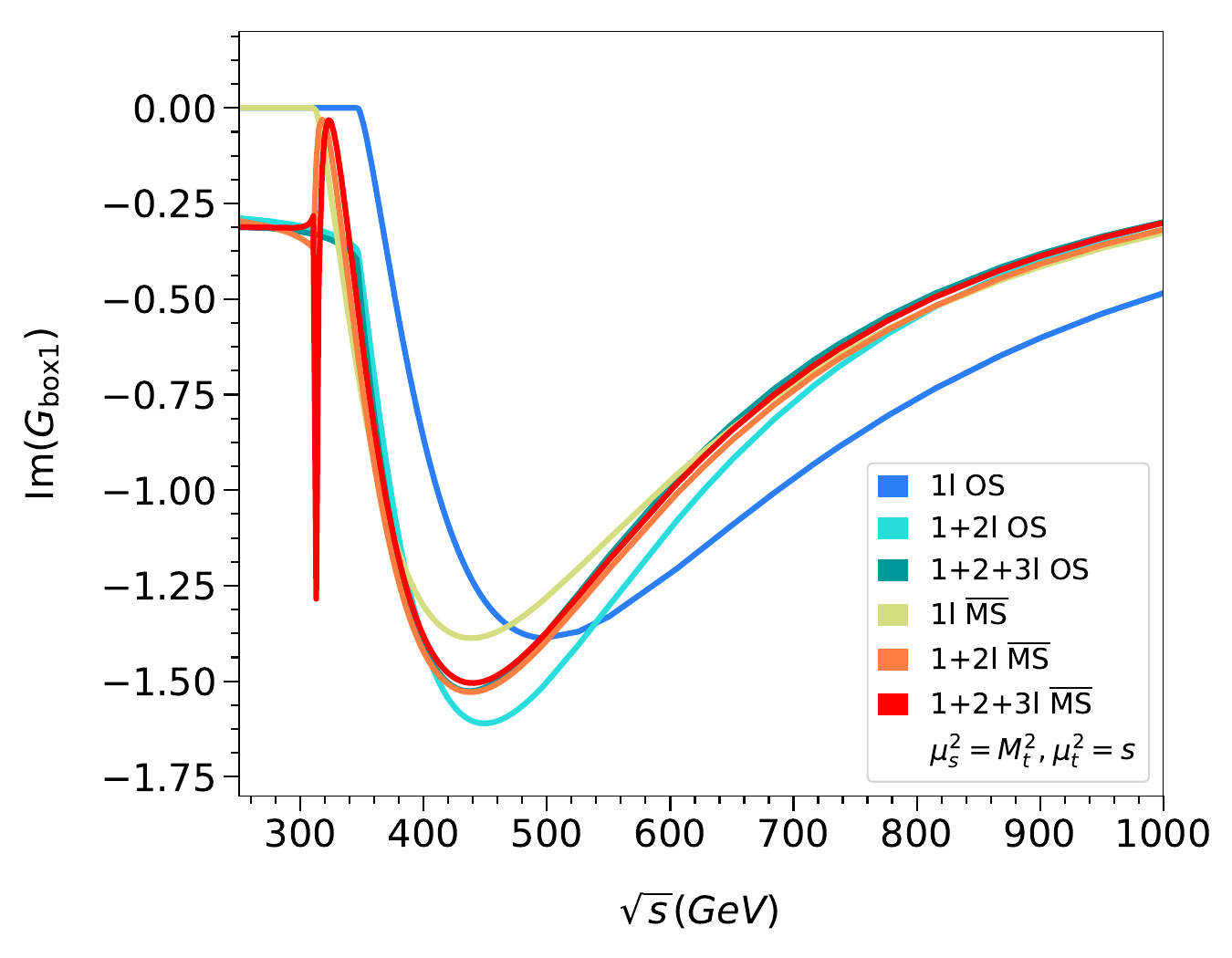}
    \\
    \includegraphics[width=0.45\textwidth]{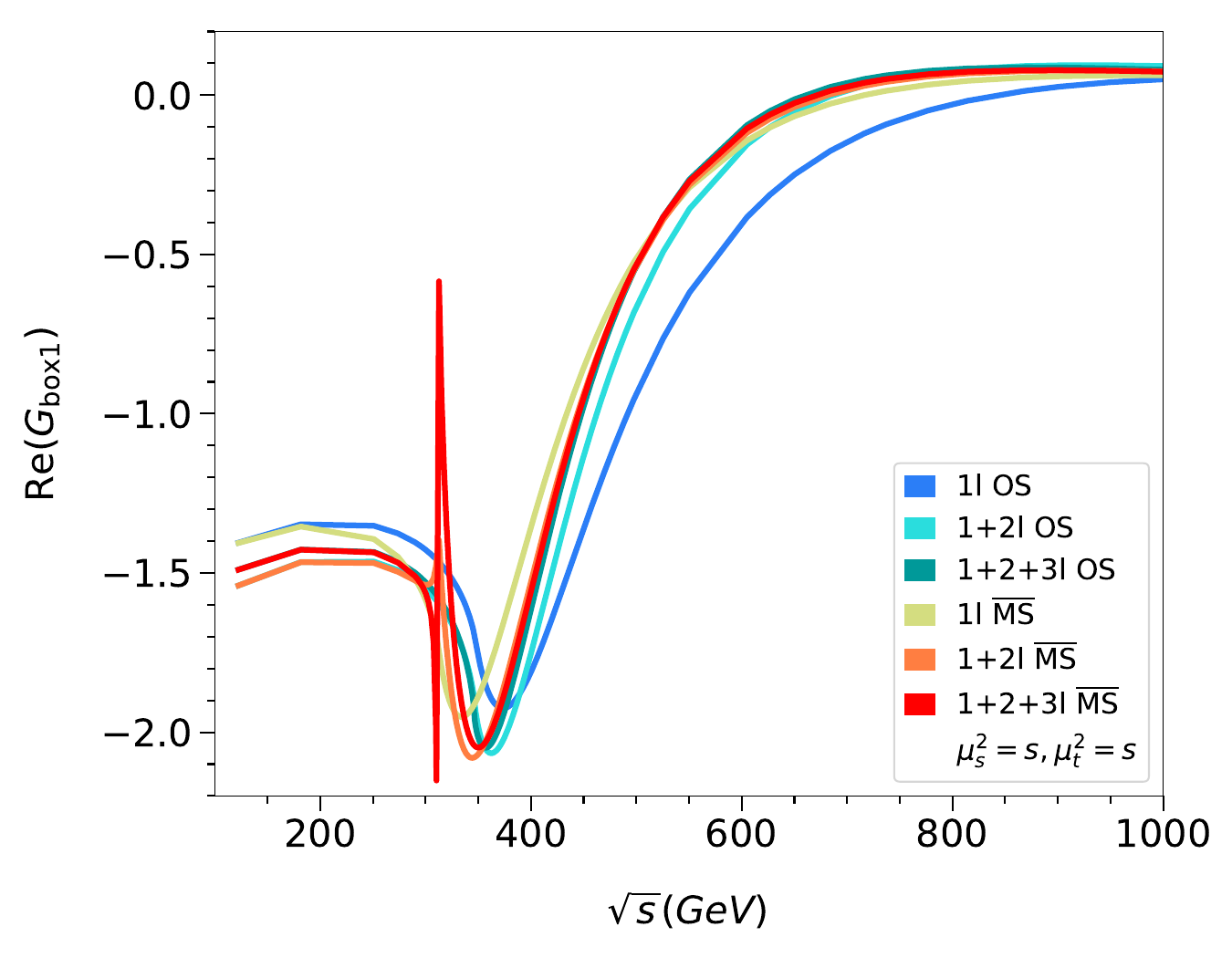} &
    \includegraphics[width=0.45\textwidth]{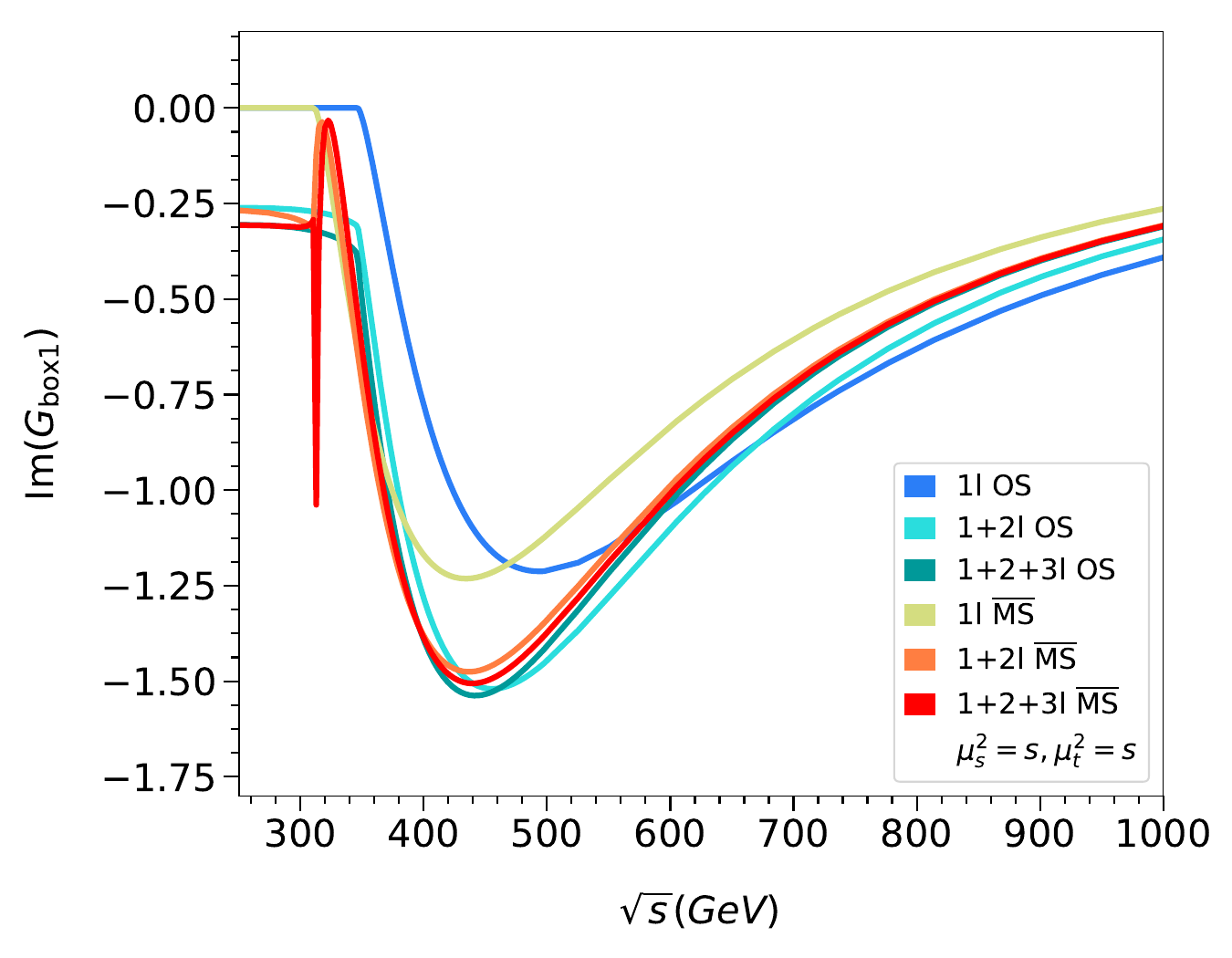}
    \\
    \includegraphics[width=0.45\textwidth]{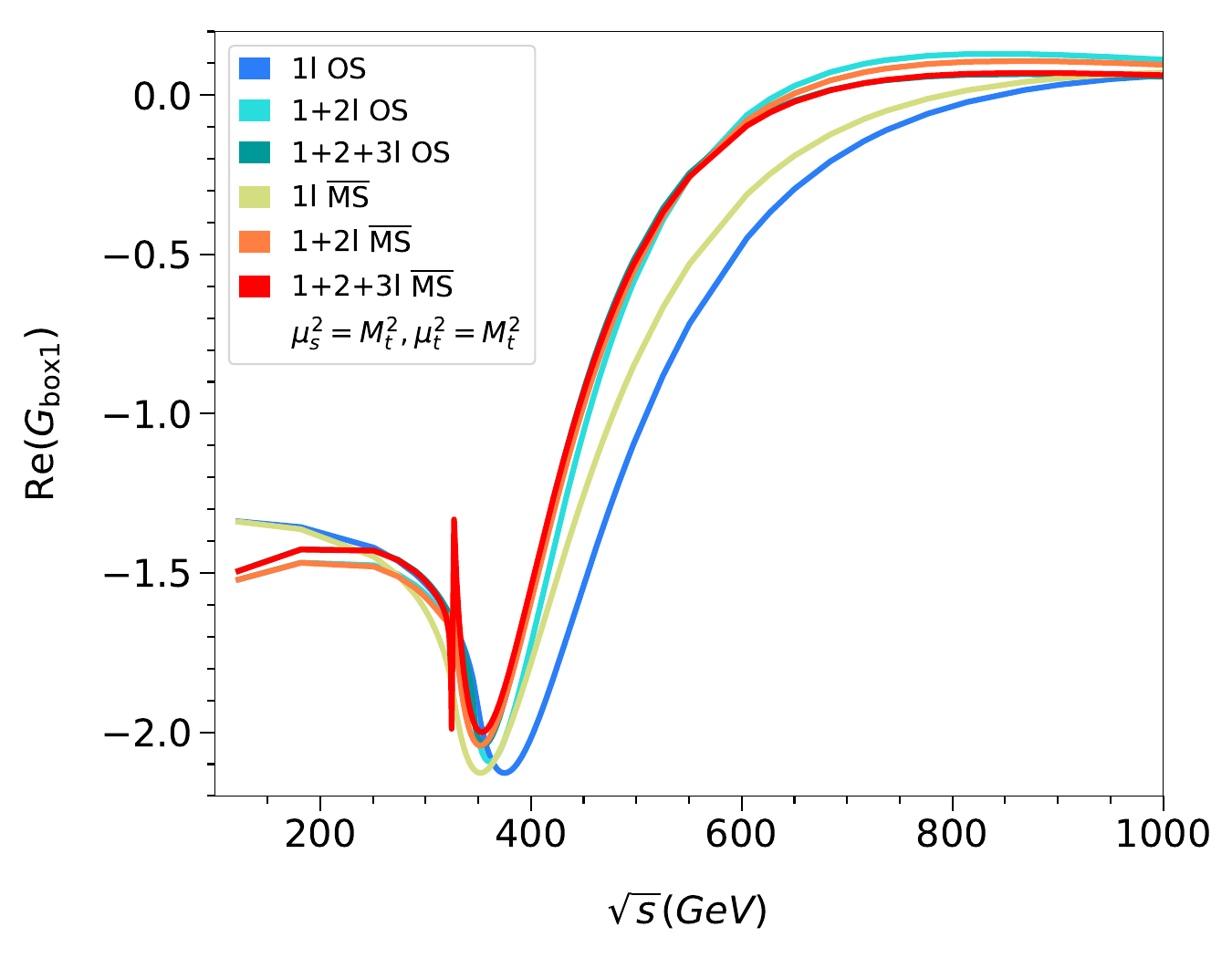} &
    \includegraphics[width=0.45\textwidth]{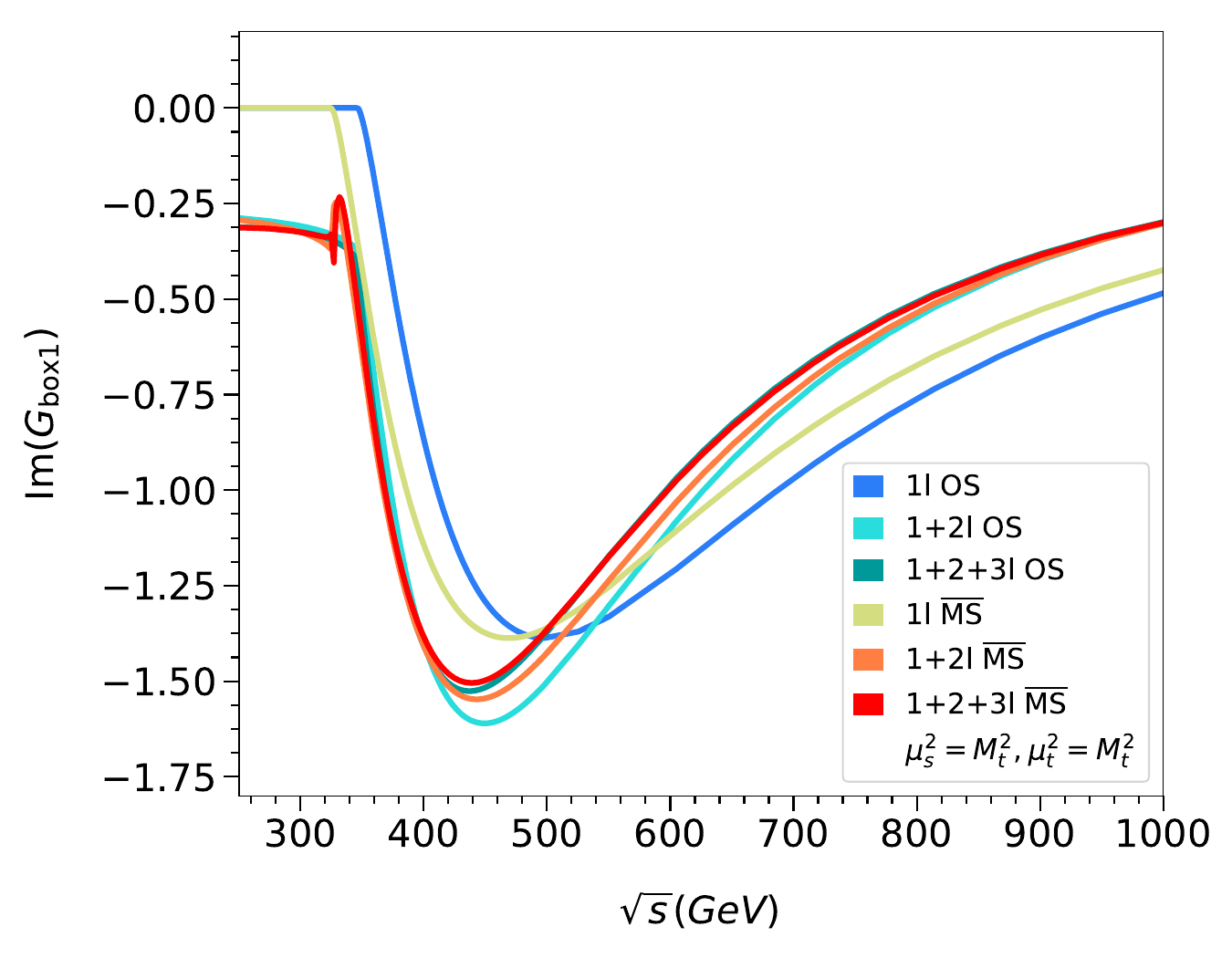}
  \end{tabular}
  \caption{\label{fig::F1_MS_OS} Real (left) and imaginary (right) parts of
    $F_1$ for $t=0$ and $m_H=0$.  At three loops the light-fermion (for
    $n_l=5$) and large-$N_c$ are included.
    Top: $\mu_s^2={M_t^2}$, $\mu_t^2=s$,
    middle: $\mu_s^2=s$, $\mu_t^2=s$,
    bottom: $\mu_s^2={M_t^2}$, $\mu_t^2={M_t^2}$
   }
\end{figure}

Let us next discuss the dependence on the top quark mass scheme at NNLO.
Fig.~\ref{fig::F1_MS_OS} shows $G_{\rm box1}$ truncated to one, two and
three loops for the pole and $\overline{\rm MS}$ schemes. The real and imaginary
parts are shown separately. In the different rows we adopt different choices for
$\mu_t$ and $\mu_s$, namely $\mu_s^2=M_t^2$, $\mu_t^2=s$ (top), $\mu_s^2=s$,
$\mu_t^2=s$ (middle) and $\mu_s^2=M_t^2$, $\mu_t^2=M_t^2$ (bottom).  
Due to
the behaviour of the $\overline{\rm MS}$ result at $s/\overline{m}_t^2=4$ we
restrict the following discussion to $\sqrt{s}\gtrsim 350$~GeV although
all curves are shown also for lower values of $\sqrt{s}$.

In all cases we observe a significant reduction on the dependence of the top
quark mass scheme when going from one to two and finally to three loops; the
one-loop curves (blue and dark yellow) are far apart. The distance is noticeably
reduced at two loops (turquoise and orange) and has almost disappeared at
three loops (green and red).  It is interesting to note that the orange and
red curves are close together which suggests that the NNLO corrections in the
$\overline{\rm MS}$ scheme are small.

\begin{figure}[t]
  \begin{tabular}{cc}
    \includegraphics[width=0.45\textwidth]{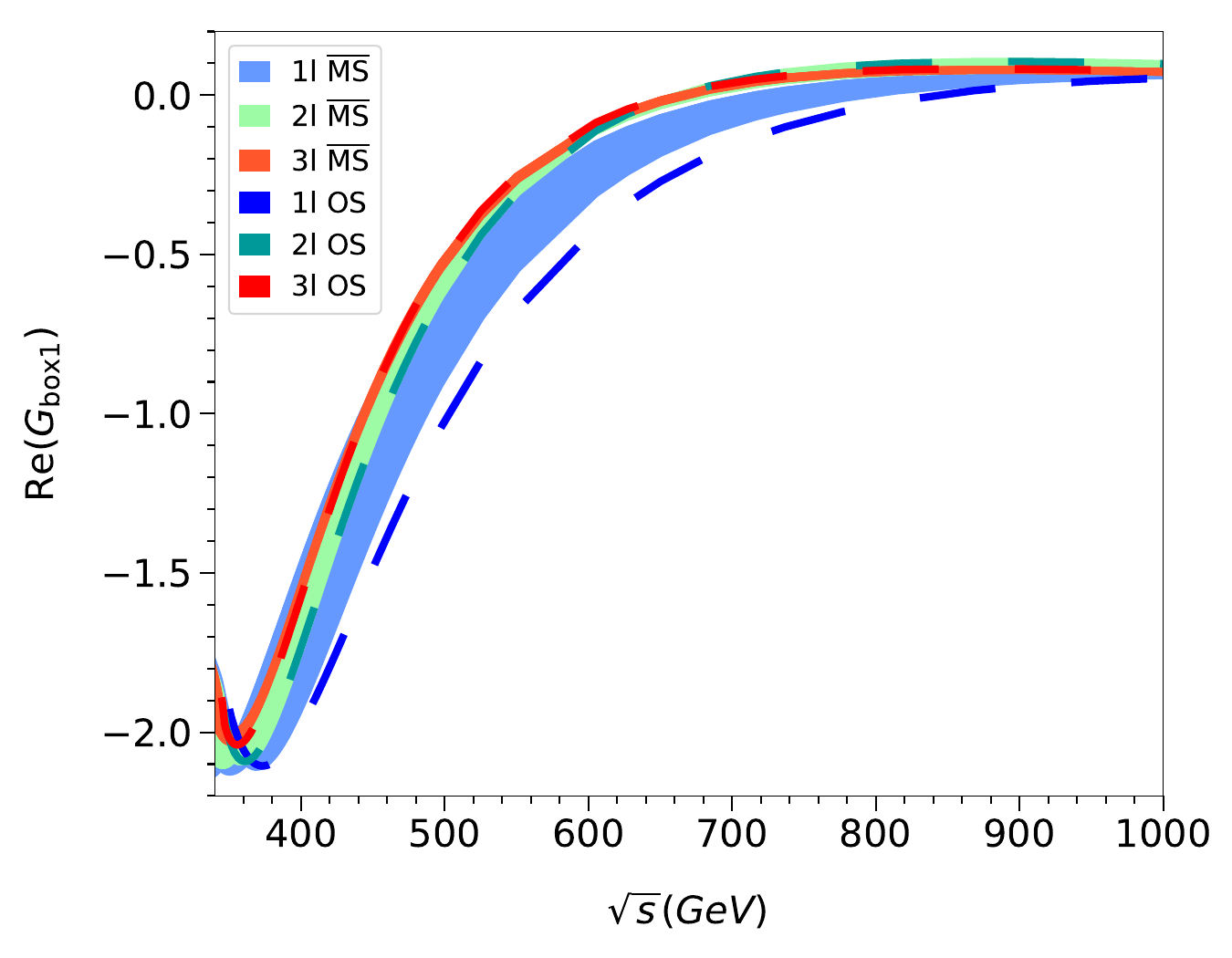} &
    \includegraphics[width=0.45\textwidth]{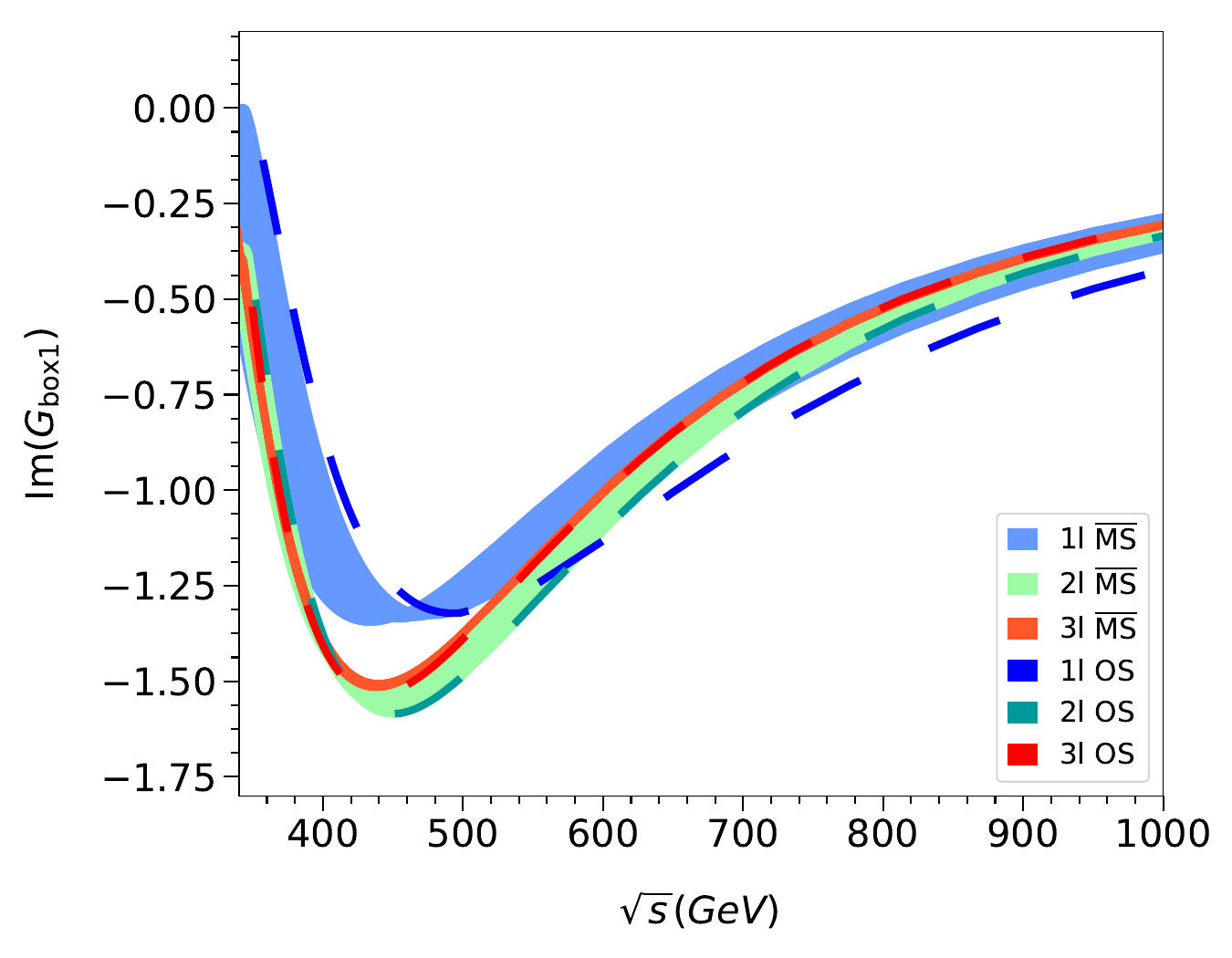}
  \end{tabular}
  \caption{\label{fig::F1_scales} $G_{\rm box1}$ for
    $\mu_s^2=s/4$. The band is the envelope of the $\overline{\rm MS}$ result where  $\mu_t^2$ is varied between $s$ and $s/16$ and $\mu_t^2={M_t^2}$ is chosen.
    The results in the pole scheme are shown as dashed lines. Note that at NNLO
    the band is quite narrow and almost completely covered by the
    curve from the pole scheme.
   }
\end{figure}

\begin{figure}[t]
\begin{center}
  \begin{tabular}{cc}
    \includegraphics[width=0.45\textwidth]{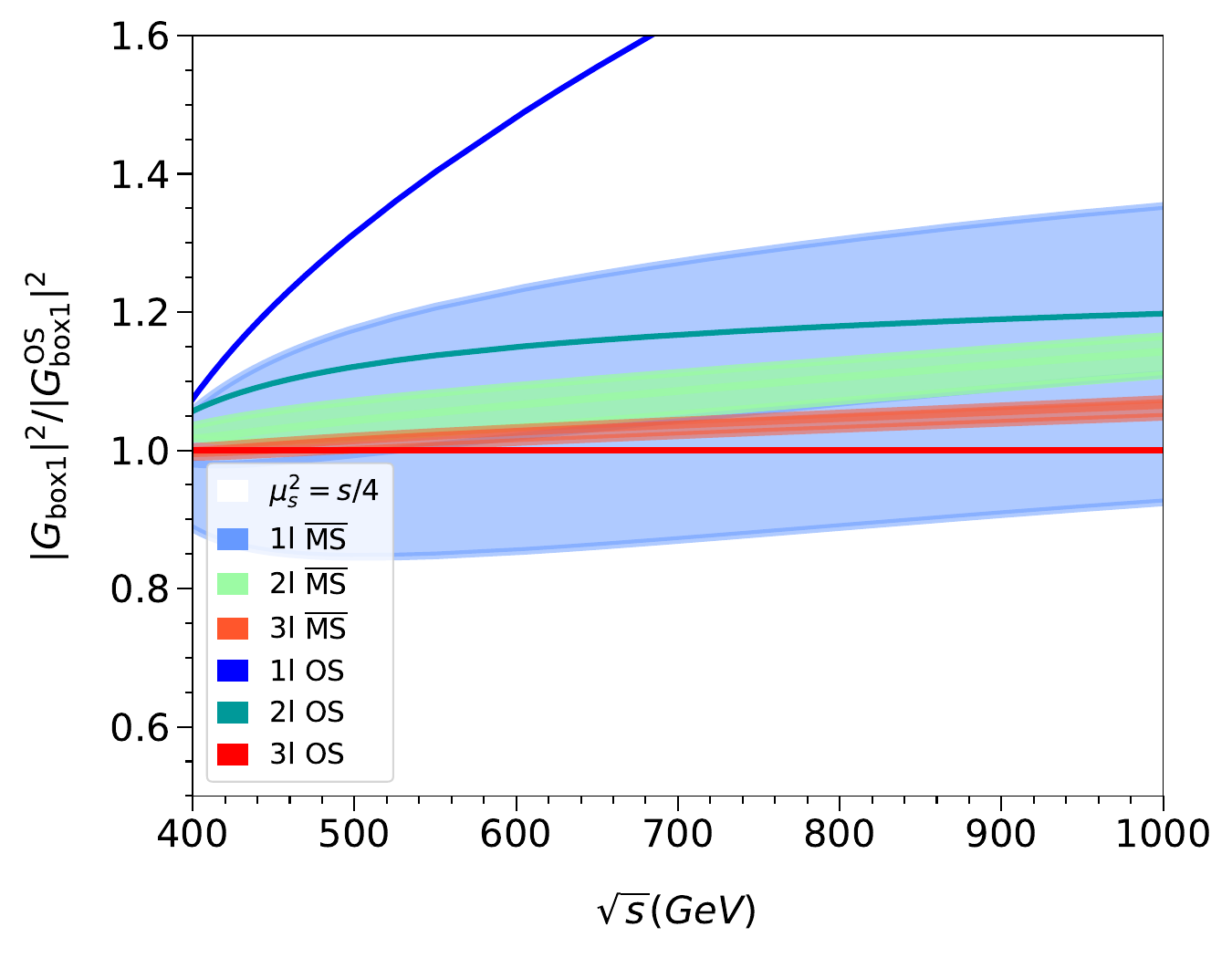}
    &
    \includegraphics[width=0.45\textwidth]{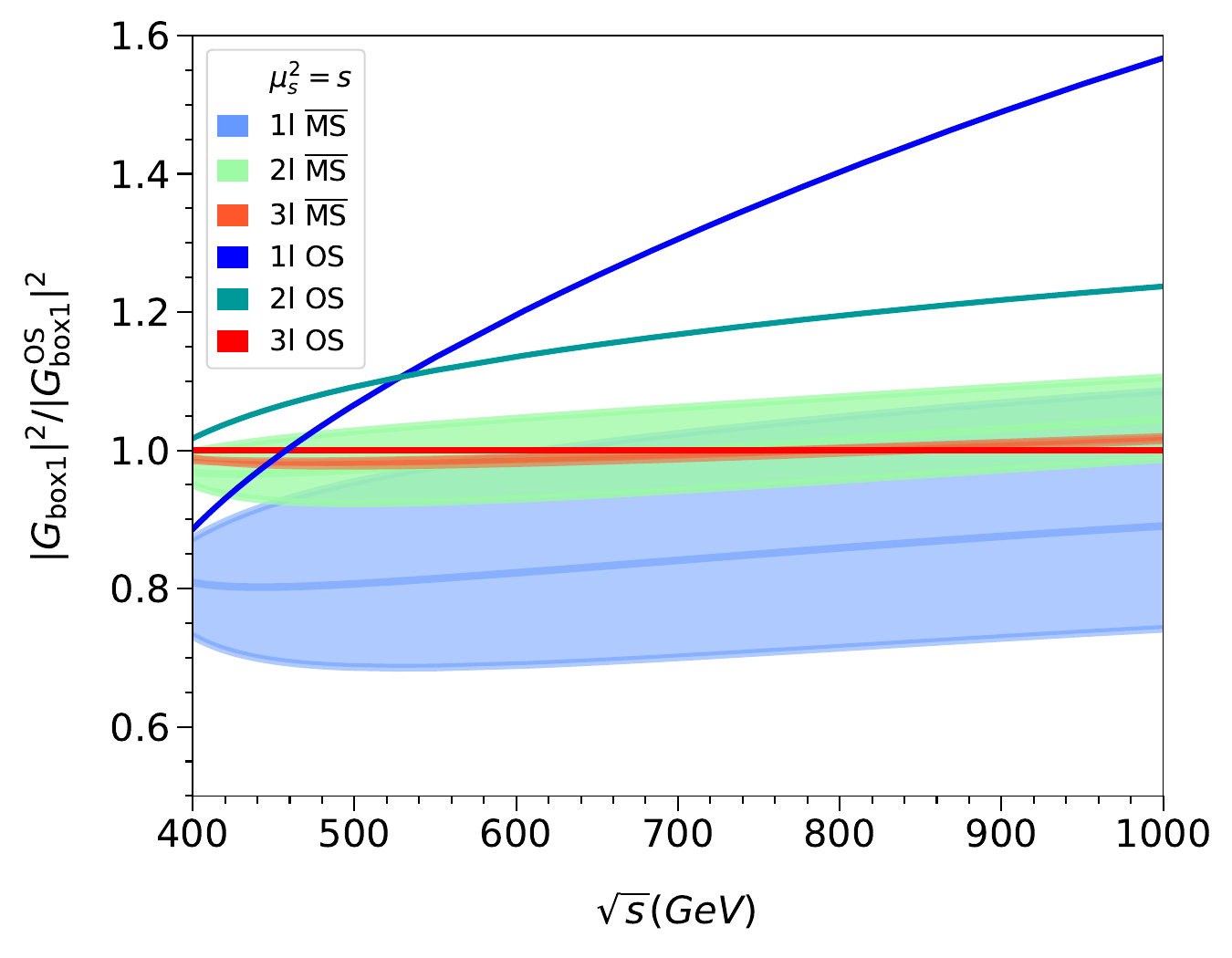}
  \end{tabular}
\end{center}
  \caption{\label{fig::G1_ratio} $G_{\rm box1}$ 
  computed with the $\overline{\rm MS}$ and on-shell
  definition of the top quark mass for 
    $\mu_s^2=s/4$ (left) and $\mu_s^2=s$ (right).  $\mu_t^2$ is chosen between  $s$ and $s/16$.  All curves are normalized to the three-loop on-shell result.}
\end{figure}

Finally, in Fig.~\ref{fig::F1_scales} we show $G_{\rm box1}$ in the pole and
$\overline{\rm MS}$ scheme for $\sqrt{s}$ values above $340$~GeV.
We choose $\mu_s^2=s/4$, and in the $\overline{\rm MS}$ scheme we vary
$\mu_t^2$ between $s$ and $s/16$. This is the partonic correspondence
to the often-used hadronic scale choice 
$\mu_s=M_{HH}/2$ and $\mu_t=M_{HH}/4,\ldots,M_{HH}$.
In Fig.~\ref{fig::F1_scales} we also include the choice
$\mu_t^2=M_t^2$.
For illustration the curves in the pole scheme are shown in a darker colour. 
The bands shown at LO, NLO and NNLO  
represent the envelope of the different choices of $\mu_t$
in the $\overline{\rm MS}$ scheme. 
The combination with the curves from the pole scheme reflect the
uncertainty due to the scheme choice for the top quark mass.

The data used in Fig.~\ref{fig::F1_scales} are also
shown in Fig.~\ref{fig::G1_ratio} after summing the squared real and imaginary parts and normalizing to the three-loop prediction in the pole scheme. In the left panel we choose $\mu_s^2=s/4$ and the blue, green and red bands are again obtained by varying $\mu_t^2$ between $s$ and $s/16$. The LO, NLO and NNLO curves
for the on-shell top quark masses are shown as dark blue,
dark green and red lines, respectively.
The reduction of the
scale dependence is clearly seen by the smaller widths of the bands when going to higher orders in perturbation theory.
Furthermore, we observe a reduction of the 
scheme dependence through the reduced distance between the 
on-shell curves and the bands which are based on $\overline{\rm MS}$ results.
We observe an overlap of the NLO and NNLO bands
for smaller values of $\sqrt{s}$ whereas for $\sqrt{s}\gtrsim 600$~GeV
there is a small gap.
Note that once also $\mu_s$ is varied 
there is an overlap of the NLO and NNLO bands. 
This can be seen in the right panel 
of Fig.~\ref{fig::G1_ratio} where $\mu_s^2=s$ has 
been chosen. We observe that the NNLO band is 
contained within the NLO scale variation.

It is interesting to note that for
smaller values for $\mu_t$
the $\overline{\rm MS}$ curves
are closer to the on-shell curves.
On the other hand, in general
for larger values of $\mu_t$
the NLO curves are more consistent with the NNLO results.




\section{\label{sec::concl}Conclusions and Outlook}

In this paper we compute the massive three-loop box-type form factors for the
process $gg\to HH$ in the large-$N_c$ limit for $p_T=0$ and massless Higgs
boson in the final state.  At one and two loops, we show that this limit
already provides a reasonable approximation to the exact results for
smaller values of $p_T$.
We additionally consider the large-$N_c$ limit, which we show to provide a
reasonable approximation in the context of the NNLO large-$m_t$ expansion.

Our calculation requires a non-trivial reduction of the box integrals to 783
master integrals. For the computation of these master integrals we apply a method which
provides semi-analytic results for the desired range of $s/m_t^2$.

We use our results to study, for the first time, the scheme dependence due to
the top quark mass at NNLO.  We find a significant reduction, as can be seen in
Fig.~\ref{fig::F1_scales} and~\ref{fig::G1_ratio}.  
If the scale uncertainty is estimated by the width of the bands in Fig.~\ref{fig::G1_ratio} we observe a typical reduction of about a factor five when going from NLO to NNLO. If we define the scheme uncertainty via the distance of the on-shell curves and the centre of the $\overline{\rm MS}$ band it  amounts to only a few percent. 
For a final phenomenological analysis it is necessary to construct physical results and include the real radiation contribution in addition.

There are a few further steps necessary before a detailed phenomenological
analysis is possible. These include the computation of the remaining colour
coefficients, the computation of sub-leading expansion terms in $t$ and $m_H$, and the
computation of the real-radiation contribution. Each step requires dedicated
technical developments and substantial computing 
resources.



\section*{Acknowledgements}  

This research was supported by the Deutsche Forschungsgemeinschaft (DFG,
German Research Foundation) under grant 396021762 --- TRR 257 ``Particle
Physics Phenomenology after the Higgs Discovery''.
The work of K.~S.~was supported by the European Research Council (ERC)
under the European Union’s Horizon 2020 research and innovation programme
grant agreement 101019620 (ERC Advanced Grant TOPUP) and the UZH Postdoc Grant,
grant no.~[FK-24-115].
The work of J.~D.~was supported by STFC Consolidated Grant ST/X000699/1.
We thank Gudrun Heinrich and Marco Vitti for carefully reading the paper and for useful comments.
The Feynman diagrams were drawn with the help of
\texttt{FeynGame}~\cite{Harlander:2020cyh,Harlander:2024qbn,Bundgen:2025utt}.



\bibliographystyle{jhep}
\bibliography{gghhnh1t0_LNC.bib}


\clearpage

\end{document}